\documentclass[preprint,12pt]{elsarticle}

\usepackage{amssymb}
\usepackage{amsthm}
\usepackage{mathrsfs}
\usepackage{graphicx}
\usepackage{amsmath}
\usepackage{amsfonts}
\usepackage{bm}
\usepackage{subcaption}
\usepackage{hyperref}
\hypersetup{
    colorlinks=true,
    linkcolor=blue,
    filecolor=magenta,      
    urlcolor=cyan,
}
\usepackage{longtable,tabularx}
\usepackage{color}
\usepackage{algorithm}
\usepackage{algpseudocode}
\usepackage{mathtools}
\usepackage{upgreek}

\newtheorem{remark}{Remark}

\newcommand{\sgn}{\mathop{\mathrm{sgn}}}

\newcommand*{\mymagenta}{\textcolor{magenta}}

\begin{document}

\begin{frontmatter}



\title{A parametric level set method with convolutional encoder-decoder network for shape optimization with fluid flow}


\author[label1]{Wrik Mallik\corref{cor1}}
\address[label1]{James Watt School of Engineering, University of Glasgow, Glasgow G12 8QQ, United Kingdom}

\cortext[cor1]{Corresponding author}

\ead{wrik.mallik@glasgow.ac.uk}

\author[label2]{Rajeev K. Jaiman}
\address[label2]{Department of Mechanical Engineering, The University of British Columbia, Vancouver, BC V6T 1Z4, Canada}

\ead{rjaiman@mech.ubc.ca}

\author[label3]{Jasmin Jelovica}
\address[label3]{Departments of Mechanical and Civil Engineering, The University of British Columbia, Vancouver, BC V6T 1Z4, Canada}

\ead{jjelovica@ubc.ca}

\begin{abstract}
In this article, we present a new data-driven shape optimization approach for implicit hydrofoil morphing via a polynomial perturbation of parametric level set representations. Without introducing any change in topology, the hydrofoil morphing is achieved by six shape design variables associated with the amplitude and shape of the perturbed displacements.  
The proposed approach has three to four times lower design variables than shape optimization via free-form deformation techniques and almost two orders lower design variables compared to topology optimization via traditional parametric level sets. 
Using the fixed uniform Cartesian level set mesh, we also integrate deep convolutional encoder-decoder networks as a surrogate of high-fidelity Reynolds-averaged Navier-Stokes (RANS) simulations for learning the flow field around morphed hydrofoil shapes. We show that an efficient shape representation via parametric level sets can enable online convolutional encoder-decoder application for the shape optimization of hydrofoils.
The generalized flow field prediction of the convolutional encoder-decoder is demonstrated by the mean structural similarity index measure (SSIM) of 0.985 and a minimum SSIM of 0.95 for the predicted solutions compared to the true solutions for out-of-training shapes.  
The convolutional encoder-decoder flow field predictions are performed nearly five orders of magnitude faster compared to their RANS counterparts. This enables a computationally tractable surrogate-based drag minimization of fifty different hydrofoil configurations for two different design lift coefficients. 
Furthermore, the best local minimum obtained via the surrogate-based optimization lie in the neighbourhood of the RANS-predicted counterparts for both the design lift cases. The present findings show promise for the future application of parametric level sets with convolutional encoder-decoder for shape optimization over a broader spectrum of flow conditions and shapes. 
\end{abstract}



\begin{keyword}
parametric level set method \sep convolutional encoder-decoder \sep shape optimization \sep hydrofoil morphing
\end{keyword}

\end{frontmatter}


\section{Introduction}
The advent of high-performance computing resources over the last three decades has led to a growing interest in shape optimization. Structural optimization studies have been increasingly focused on shape and topology, whereas past studies relied solely on sizing optimization \cite{haftka1986structural}. Similarly, accurate computation of shape sensitivities of even complex configurations in fluid flows \cite{jameson_1998,Giles_2000,pironneau_2004} has led to a growing interest in aerodynamic shape optimization over the last two decades \cite{reuther1996aerodynamic,samareh2004aerodynamic}. However, integrating high-fidelity computational structural mechanics or computational fluid dynamics (CFD) simulations within an optimization algorithm is often restricted by the huge computational expense of these simulations. Apart from the function evaluations via these high-fidelity solvers, the cost of numerical sensitivity computations can be large even when adjoint-based techniques are employed \cite{jameson_1998,samareh2004aerodynamic}. Some shape optimization studies with the fluid flow have recently been performed with high-fidelity CFD solvers \cite{garg2015high,lyu2015aerodynamic,he2019robust} with powerful computing resources at a great computational cost. These studies relied on a gradient-based search algorithm with either a single or very few initial conditions, and cannot explore the full design space owing to their computational cost. However, the presence of local optima and multimodality in aerodynamic design are well-documented \cite{chernukhin2013multimodality} and computational efficiency must be considered in this regard.

Various data-driven model reduction and surrogate modelling techniques have been developed over the last few years to reduce the online computational cost for shape optimization. Kriging has been a widely used surrogate modelling method in various engineering analysis and design tasks \cite{timme2011transonic,fan2019reliability,mallik2020kriging}. However, Kriging suffers from a variety of issues such as being poor at approximating discontinuous functions \cite{raissi2016deep}, difficulty in handling
high-dimensional problems \cite{perdikaris2017nonlinear}, expensive to use in the presence of a large number of data samples \cite{keane2008engineering}, and is difficult to implement for solving certain inverse problems with strong nonlinearities \cite{bonfiglio2018improving}. Recently, reduced-order models obtained via proper orthogonal decomposition \cite{yao2020reduced}, and hyper-reduction based on proper orthogonal decomposition modes \cite{marques2021non}, have also been employed. More recently, deep learning models based on deep neural network architectures have found widespread application as surrogate models. Most of these models have been developed via multi-layer perceptron or the standard feedforward neural network architecture \cite{bouhlel2020scalable, junior2022intelligent}. Such feedforward neural networks obtain a map directly from the design parameters to the optimization objective and constraint outputs via nonlinear regression. However, the surrogate models obtained with multi-layer perceptron networks cannot retrieve the flow field around the morphed configurations. Thus, the feedforward neural networks behave completely like a ``black box" surrogate model and lack interpretability.

Recently, convolutional neural networks (CNNs) have been employed to learn the flow field around airfoils at various steady flow conditions \cite{bhatnagar2019prediction} and the unsteady flow around bluff bodies with a fixed fluid-solid interface \cite{xu2020multi}. CNNs have been architecturally developed with various translational invariance and equivariance properties and are better at generalized learning compared to fully regression-based models \cite{bronstein2017geometric}. Also, since they do not behave completely as a nonlinear regression model, they provide better physical interpretability \cite{mallik2022assessment}. Thus, they could also potentially learn the flow field around the changing fluid-solid interface during a shape optimization or morphing operation.

The main challenge of including CNNs as a surrogate model in shape optimization is that CNNs require shape representation on a fixed uniform Cartesian grid \cite{bhatnagar2019prediction}. Thus, they do not complement the free-form deformation (FFD) technique \cite{sederberg1986free}, or any of its popular modifications \cite{kenway2010cad,garg2015high,marques2021non,junior2022intelligent}, which employ a moving grid. A recent study in aerodynamic shape optimization has proposed the application of level set methods for shape representation to include CNNs as a surrogate model for aerodynamic force coefficient prediction \cite{mallik2022deep}. Such level set methods enable implicit representation of any general complex shape or topology on a fixed Cartesian grid and are frequently observed in the topology optimization literature \cite{sethian1996theory, allaire2002level,wang2006radial,wang2007extended,jiang2018parametric}. Various parametric level set methods have since been developed via radial basis functions (RBF) \cite{wang2006radial, wang2007extended, aghasi2011parametric,pingen2010parametric,jiang2018parametric} for numerically stable optimization of complex topologies. However, in these RBF-based parametric level sets a single topology design parameter is associated with each level set grid point over the whole domain, leading to a very large number of optimization design variables. If such a parametric level set implementation is also considered for shape optimization without any topology change, the design variables will easily outnumber their counterparts from routine FFD-discretization, which only parameterize the boundary and not the full domain. Thus, unless more efficient employment of parametric level set methods is developed for shape optimization, the computational cost and complexity of employing CNNs as a surrogate model in shape optimization far outweigh any benefit over present feedforward neural network-based surrogates \cite{bouhlel2020scalable,junior2022intelligent}.

\subsection{Contributions}
In this article, we propose a new approach of employing the RBF-based parametric level set method for morphing generic bluff or streamlined body shapes. We specifically consider the parametric RBF-based level sets with polynomials \cite{wang2006radial, wang2007extended}. For topology optimization with these parametric level sets, the standard approach is to consider RBF coefficients associated with each level set grid points as topology design variables \cite{wang2006radial, wang2007extended}. We show that for shape optimization, we only need to modify the polynomial coefficients of the parametric level sets in a prescribed manner to obtain morphed shapes, without modifying any of the RBF coefficients. The present approach enables us to perform shape morphing with only six design variables, which is almost two orders reduction in design variables compared to the conventional use of RBF-based parametric level sets in optimization \cite{wang2006radial, wang2007extended, pingen2010parametric,jiang2018parametric,guirguis2018high}. Furthermore, it also enables smooth shape change without any change in topology. At the same time, the implicit nature of the parametric level set is retained so that generic hydrofoil shapes can be achieved during the shape optimization with this approach. This is in contrast to a coarse airfoil grid discretization using FFD, which will have limited shape sensitivity accuracy, but still ends up requiring twenty design parameters \cite{junior2022intelligent}. 

Finally, with the direct application of a computationally efficient level set method for shape representation and optimization, we can seamlessly integrate a CNN-based surrogate model for real-time flow field prediction around the morphed configurations. CNNs have shown an excellent capacity for learning fluid flow physics at various steady \cite{bhatnagar2019prediction} and unsteady flow conditions \cite{xu2020multi}. However, their capacity to learn flow physics around deforming fluid-solid boundaries and their online application for shape optimization is yet to be demonstrated. Thus, such surrogate modelling application of CNNs for shape optimization is also presented here for the first time.

The research presented here can potentially impact the aerospace ad ocean engineering industry by optimally morphing the shape design of aerofoils and hydrofoils. The urgent need for online shape optimization of marine propellers to deal with cavitation and underwater radiated noise is already documented \cite{vernengo2016physics, gaggero2017efficient, miglianti2020predicting}. Thus, the present CNN-based shape optimization approach can be a promising avenue for real-time marine propeller optimization over a spectrum of hydrofoil shapes and underwater flow conditions.



\subsection{Organization}
The paper is structured as follows. Section 2 explains the novel application of the RBF-based parametric level set for shape morphing. Both its application during the offline phase for generating CFD mesh of morphed configurations and during the online phase for shape optimization are discussed. In section 3, we explain the development of the convolutional encoder-decoder, a fully CNN-based surrogate model of the high-fidelity Reynolds-averaged Navier-Stokes (RANS) simulations. The non-intrusive gradient-based shape optimization problem is formulated in section 4. In section 5 we present how our complete methodology of CNN-based 
shape optimization can be employed for a test problem. We specifically select the NACA66 hydrofoil, which is routinely employed as marine propeller blade sections. In section 6, we present the performance of the convolutional encoder-decoder to predict the flow field for various morphed configurations outside the training set and the optimization results with the convolutional encoder-decoder as the surrogate model. The last section concludes the study presented here.

\section{Parametric level-set functions}
Level set methods are well-known for their capacity to implicitly represent even complex shapes and topologies \cite{sethian1996theory}. The geometry is represented implicitly on a Cartesian grid via the widely used signed distance functions \cite{sethian1996theory,sethian1999level,allaire2002level}. Mathematically, a signed distance function $f\left(\bm{\mathrm{x}}\right)$, of a set of points $\bm{\mathrm{X}}\subset \mathbb{R}^d \left(d=1,2,3\right)$ determines the minimum distance of each
given point $\bm{\mathrm{x}}\in \bm{\mathrm{X}}$ from the boundary of an object. Thus
\begin{equation}
    \label{eq:level-set_def}
    f\left(\bm{\mathrm{x}}\right) = \begin{cases}
    \min_{\bm{\mathrm{x}}_I \in \partial\Omega} \left({\Vert \bm{\mathrm{x}} - \bm{\mathrm{x}}_I \Vert}\right), &\quad \bm{\mathrm{x}} \notin \Omega\\
    0, &\quad \bm{\mathrm{x}} \in \partial\Omega\\
    -\min_{\bm{\mathrm{x}}_I \in \partial\Omega} \left({\Vert \bm{\mathrm{x}} - \bm{\mathrm{x}}_I \Vert}\right), &\quad \bm{\mathrm{x}} \in \Omega\\
    \end{cases}
    ,
\end{equation}
where $\partial \Omega$ represents the shape boundary and $\lVert \cdot \rVert$ denotes the Euclidean norm. The distance sign determines whether the given point is inside or outside of the object. 

Direct application of level set methods \cite{sethian1999level} can pose numerical challenges for tracking the shape or topology boundary during level set evolution \cite{wang2006radial}. Furthermore, they are not amenable to generic optimization algorithms. Thus, various parametric level set methods have been developed with radially symmetric basis functions for describing the level set \cite{wang2006radial,wang2007extended,pingen2010parametric,jiang2018parametric}. Such parametric representation with RBFs enables easy interpolation of the boundary throughout the domain and also easy integration with routine optimization algorithms. The parametric level set with RBFs is often further supported by first-degree polynomials to ensure positive definiteness of the level set function \cite{wang2006radial}. RBF-based parametric level set with a polynomial \cite{wang2006radial,wang2007extended}, which can be written as,
\begin{equation}
    \label{eq:parametric_level_set_polynomial}
    \Phi(\bm{\mathrm{x}}) = \sum_{i=1}^{N_c}\beta_i \phi \left(\lVert \bm{\mathrm{x}} - \bm{\mathrm{x}}_i^c \rVert\right) + p(\bm{\mathrm{x}}).
\end{equation}
where $\phi$ is the set of the radial basis functions, $\beta_i$ are the weights of the basis functions and the first-degree polynomial $p(\bm{\mathrm{x}}) = \lambda_0 + \lambda_1 x + \lambda_2 y + \lambda_3 z$. To obtain unique solutions for the coefficients $\beta_i$ we must satisfy
\begin{equation}
    \Phi(\bm{\mathrm{x}}_i^c)=f(\bm{\mathrm{x}}_i^c),1\leq i\leq N_c
\end{equation}
and $\phi\left(\lVert \cdot \rVert\right)$ should be positive definite in nature. $\bm{\mathrm{X}}^c = \left\{\bm{\mathrm{x}}_1^c,\bm{\mathrm{x}}_2^c,\ldots,\bm{\mathrm{x}}_{N_c}^c\right\}\subseteq \mathbb{R}^d$ and denotes the set of control points on a fixed Cartesian grid. 

Traditional topology optimization with parametric level sets, both with \cite{wang2006radial,wang2007extended} or without polynomials \cite{pingen2010parametric,jiang2018parametric}, assign the $N_c$ RBF kernel weights as design variables. For a Cartesian grid in two-dimensions with $n$ and $m$ discretization along the $x$ and $y$ axes respectively, $N_c=(m+1)\times (n+1)$. Thus, even for a coarse level set grid, this can be a large number. Furthermore, the topology optimization design variables will increases exponentially with the spatial dimension, resulting in huge computational costs for the optimization. 

Here we will use parametric level set methods for shape optimization without causing any topology change. We will specifically use the parametric level set representation with polynomials. The level set method presented here will significantly reduce the number of design variables required for level set-based optimization. This will make them competitive or even better than popularly used FFD-based shape optimization, in terms of computational expense. Furthermore, a smooth change in shape without any change in topology is difficult to control via a direct change in RBF kernel function weights. We will also show how our proposed application of the parametric level set methods can solve this problem. Another challenge for level set representations is to retrieve the boundary ($\Phi'(\bm{\mathrm{x}})=0, \bm{\mathrm{x}} \in \partial \Omega'$) for each morphed configuration with sufficient accuracy such that we can generate a well-conditioned mesh to numerically solve the governing equations for each configuration. In the following subsections, we will also explain how we solve this challenge with our level set-based approach.

 


\subsection{Level set function perturbation with minimal parameters}
In the level set-based shape optimization we want to perturb our initial shape representation our $\Phi(\bm{\mathrm{x}})$. This is achieved by directly perturbing the original shape boundary $\Phi(\bm{\mathrm{x}})=0, \bm{\mathrm{x}} \in \partial \Omega$ to the perturbed shape boundary $\Phi'(\bm{\mathrm{x}})=0, \bm{\mathrm{x}} \in \partial \Omega'$. We propose that this be achieved smoothly during morphing/optimization, without any change in topology, by only modifying the polynomials of the parametric level set. The RBF coefficients are left as same as that of the initial/baseline configuration ($\Phi(\bm{\mathrm{x}})$). 

\begin{remark}
For polynomials of the form $p(\bm{\mathrm{x}}) = \lambda_0 + \lambda_1 x + \lambda_2 y + \lambda_3 z$, the shape boundary ($\Phi(\bm{\mathrm{x}})=0$) can be stretched (or compressed) uniformly, proportional to $\lambda_0$. Also, the boundary can be stretched (or compressed) linearly along the $x$, $y$ and $z$-axes proportional to $\lambda_1$, $\lambda_2$ and $\lambda_2$, respectively. However, such morphing is achieved without any topology change of the configuration i.e., no new holes or boundaries are created. Thus, perturbation of $p(\bm{\mathrm{x}})$ enables morphing into a wide range of shapes with smooth boundaries. 
\end{remark}

\begin{figure}[ht]
\centering
    \begin{subfigure}[b]{.485\textwidth}
        \centering
        \includegraphics[width=\textwidth]{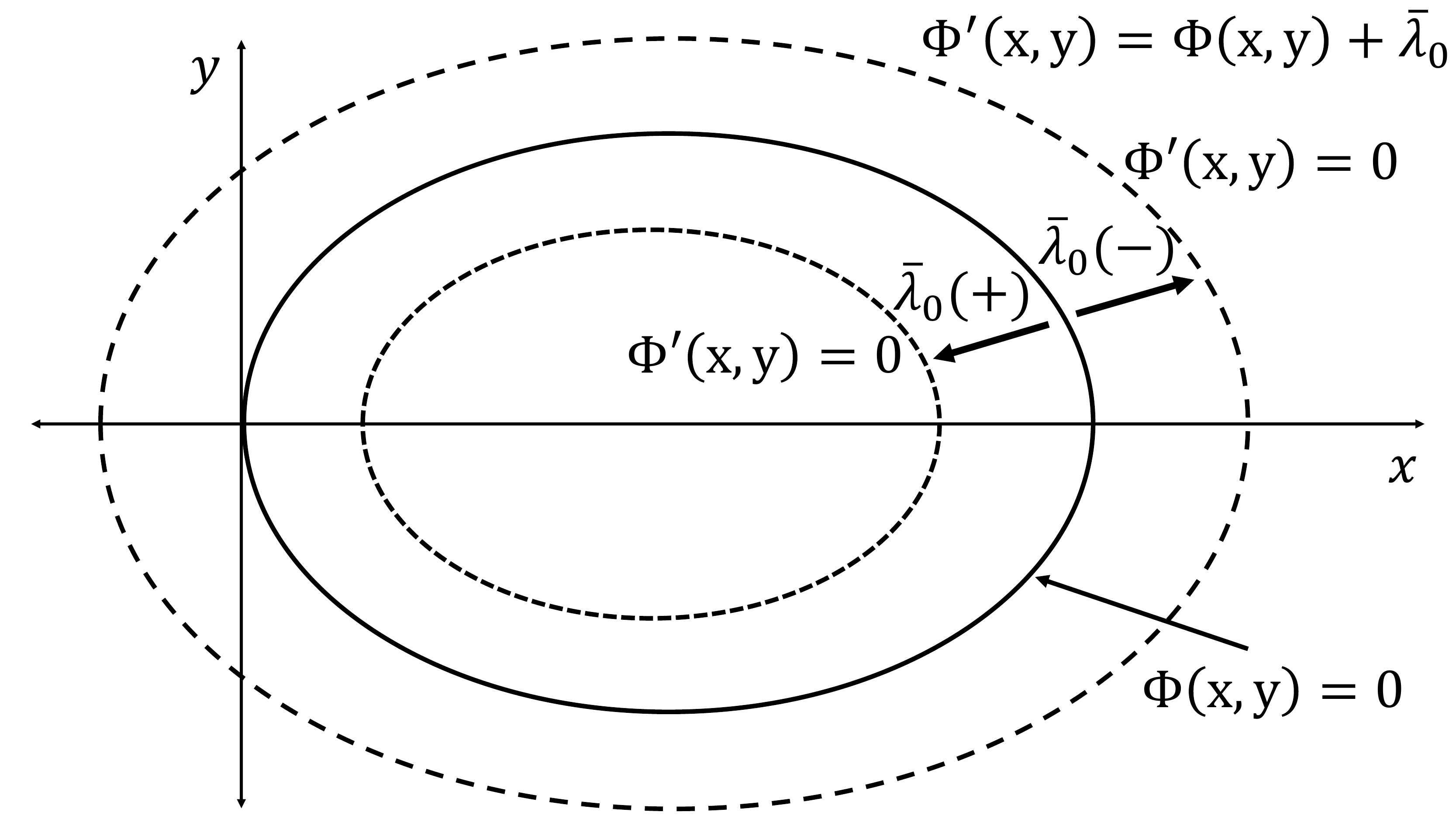}
        \caption{Uniform stretching or compressing}
    \end{subfigure}
    \hfill
    \begin{subfigure}[b]{.485\textwidth}
        \centering
        \includegraphics[width=\textwidth]{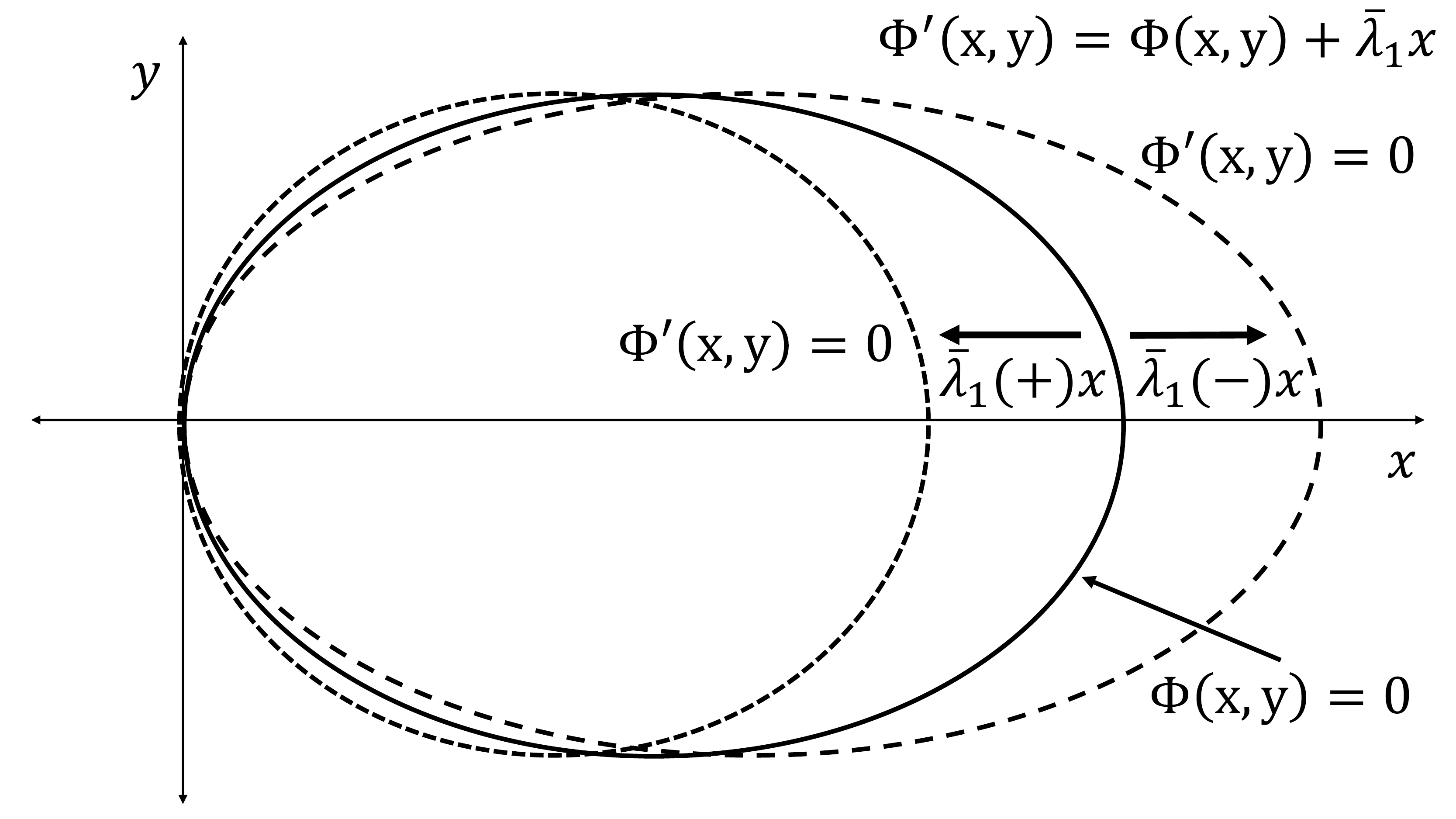}
        \caption{Perturbation along the $x$ axis}
    \end{subfigure}
    \hfill
    \begin{subfigure}[b]{.485\textwidth}
    \centering
    \includegraphics[width=\textwidth]{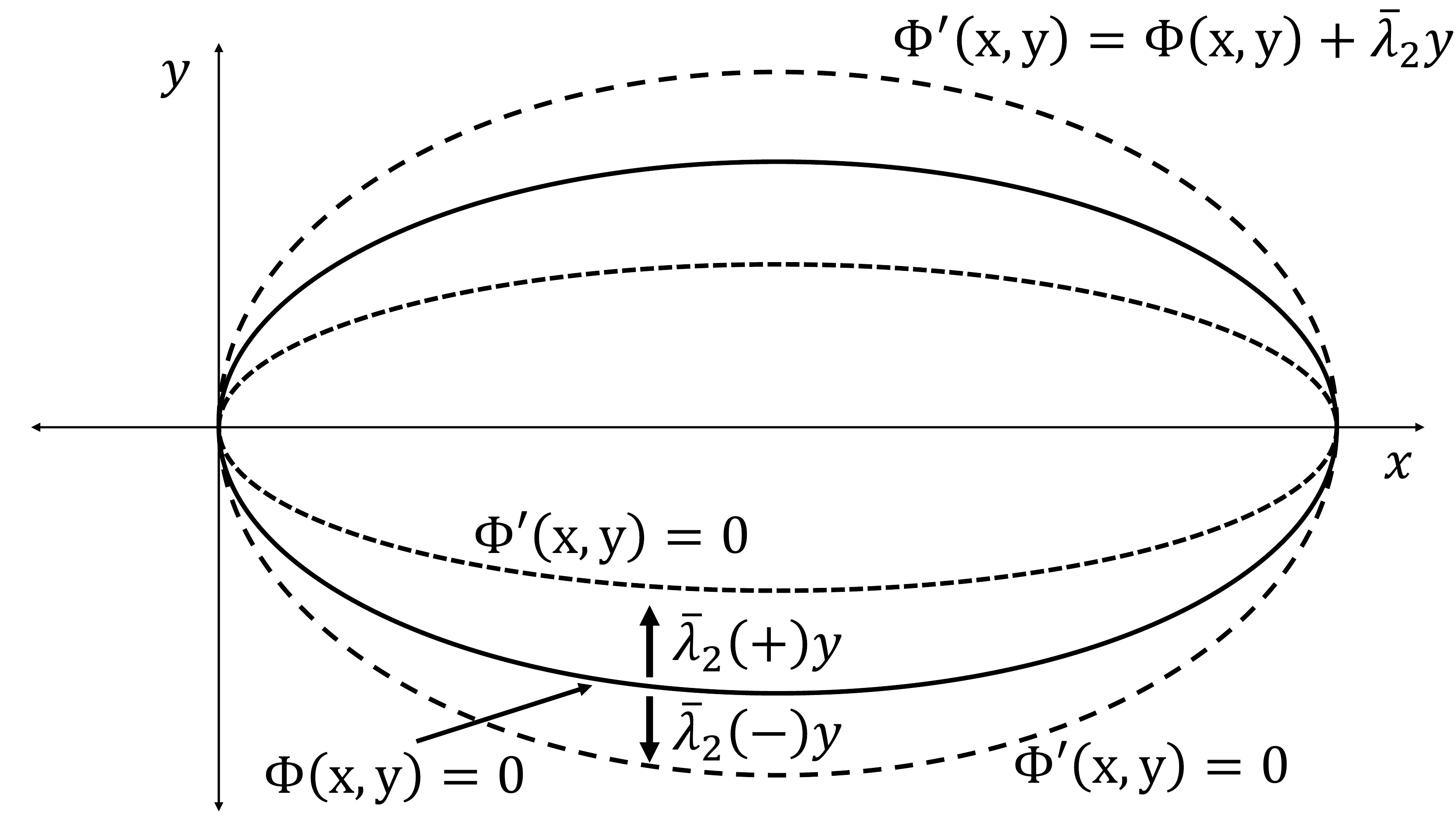}
         \caption{Perturbation along the $y$ axis}
    \end{subfigure}
\caption{An illustration of the perturbation parameters to morph two-dimensional shape}
\label{fig:param_shape_perturb}
\end{figure}

The perturbed parametric level set function $\Phi'(\bm{\mathrm{x}})$ for any general polynomial perturbation $p'(\bm{\mathrm{x}})$ is obtained as
\begin{equation}
\begin{split}
    \label{eq:perturb_approach}
    \Phi'(\bm{\mathrm{x}}) &= \Phi(\bm{\mathrm{x}}) + p'(\bm{\mathrm{x}})\\
    &= \sum_{i=1}^{N_c}\beta_i \phi \left(\lVert \bm{\mathrm{x}} - \bm{\mathrm{x}}_i^c \rVert\right) + p(\bm{\mathrm{x}}) + p'\bm{\mathrm{x}}.
\end{split}
\end{equation}
In order to perturb the boundary along the $y$-axis we can simply substitute $p'(\bm{\mathrm{x}})$ in Eq. \ref{eq:perturb_approach} with $\Bar{\lambda_2} y$ without any loss in generality. We could also perturb the parametric level set function with $\Bar{\lambda}_0$, $\Bar{\lambda}_1 x$ or $\Bar{\lambda}_1 z$, or a linear superposition of such perturbation to obtain complex morphing patterns. Fig. \ref{fig:param_shape_perturb} (a) shows how such perturbations morph the boundaries of an elliptical shape via uniform stretching or compression. Figs. \ref{fig:param_shape_perturb} (b) and (c) shows how the perturbations morph the boundaries of an elliptical shape via stretching or compression along the $x$ and $y$-axes, respectively. Similarly, we can also apply the polynomial perturbation approach to morph any general three-dimensional shape.

The proposed morphing approach is implemented as follows by starting with the parametric level set function $\Phi(\bm{\mathrm{x}})$ of our baseline/initial shape. In order to obtain unique RBF and polynomial coefficients, $\Phi$ must not only match the level set values at the control points ($\Phi(\bm{\mathrm{x}}_i^c)=f_i^c,1\leq i\leq N_c$)  but the following constraints must also be satisfied,
\begin{equation}
    \label{eq:add_constrnts}
    \sum_{i=1}^{N_c} \beta_i=0, \quad \sum_{i=1}^{N_c} \beta_i x_i=0, \quad \sum_{i=1}^{N_c} \beta_i y_i=0.
\end{equation}
This leads us to the following set of linear equations
\begin{equation}
    \bm{K} \bm{\Lambda} = \bm{f}^c,
\end{equation}
where 
\begin{align}
    \label{eq:linear_sdf_eq_coeffs}
    \bm{K} &= 
    \begin{bmatrix}
    \bm{0} &\bm{P}^T\\
    \bm{P} & \bm{A}
    \end{bmatrix},\\
    \bm{\Lambda} &= {\left[\lambda_0 \; \lambda_1 \; \lambda_2\; \beta_1\;\ldots\;\beta_{N_c}\right]}^T,\\
    \bm{f}^c &= {\left[0\; 0 \; 0\; f_1^c \; \ldots \; f_{N_c}^c\right]}^T,\\
    \bm{P} &=
    \begin{bmatrix}
    1 & x_1^c & y_1^c \\
    \vdots & \vdots & \vdots \\
    1 & x_{N_c}^c & y_{N_c}^c
    \end{bmatrix},\\
    A_{i,j} &= \phi \left(\lVert \bm{\mathrm{x}}_j^c - \bm{\mathrm{x}}_i^c \rVert\right).
\end{align}
Therefore, the coefficients of the RBF and the polynomial can be calculated by solving $\bm{\Lambda} = \bm{K}^{-1} \bm{f}^c$. We obtain our perturbed level set function for a prescribed set of parameter values following Eq. \ref{eq:perturb_approach}.

Various choices of RBFs have been popularly implemented, like Gaussian, multiquadric and polyharmonic, etc. Here we employ Wendland's $C^2$ functions \cite{wendland1995piecewise}, which not only have positive definiteness property, it is also beneficial for the radial basis function to have localization or compact support. Such localization ensures that the system matrix is sparse, which is helpful in its inversion. They are of the form
\begin{equation}
    \phi\left(r\right) = \begin{cases}
    p(r), &\quad 0 \leq r \leq 1\\
    0 &\quad r > 1 \\
    \end{cases}
    ,
\end{equation}
where $p(r)$ is a univariate polynomial. Following Wendland's recommendation for two-dimensional problems \cite{wendland1995piecewise}, we use the following function 
\begin{equation}
   \label{eq:Wendland's_2D_basis} \phi\left(\lVert\bm{\mathrm{x}}\rVert/\rho\right) = \begin{cases}
    \left(1 - \lVert\bm{\mathrm{x}}\rVert/\rho\right)^4 \left(1 + 4 \lVert\bm{\mathrm{x}}\rVert/\rho\right), &\quad 0 \leq \lVert\bm{\mathrm{x}}\rVert/\rho \leq 1\\
    0 &\quad \lVert\bm{\mathrm{x}}\rVert/\rho > 1 \\
    \end{cases}
    .
\end{equation}
Here $\rho$ represents the compact support radius, selected according to the prescribed bounds of $\bm{\mathrm{x}}$.

\begin{figure}[ht]
\centering
\includegraphics[width=0.96\textwidth]{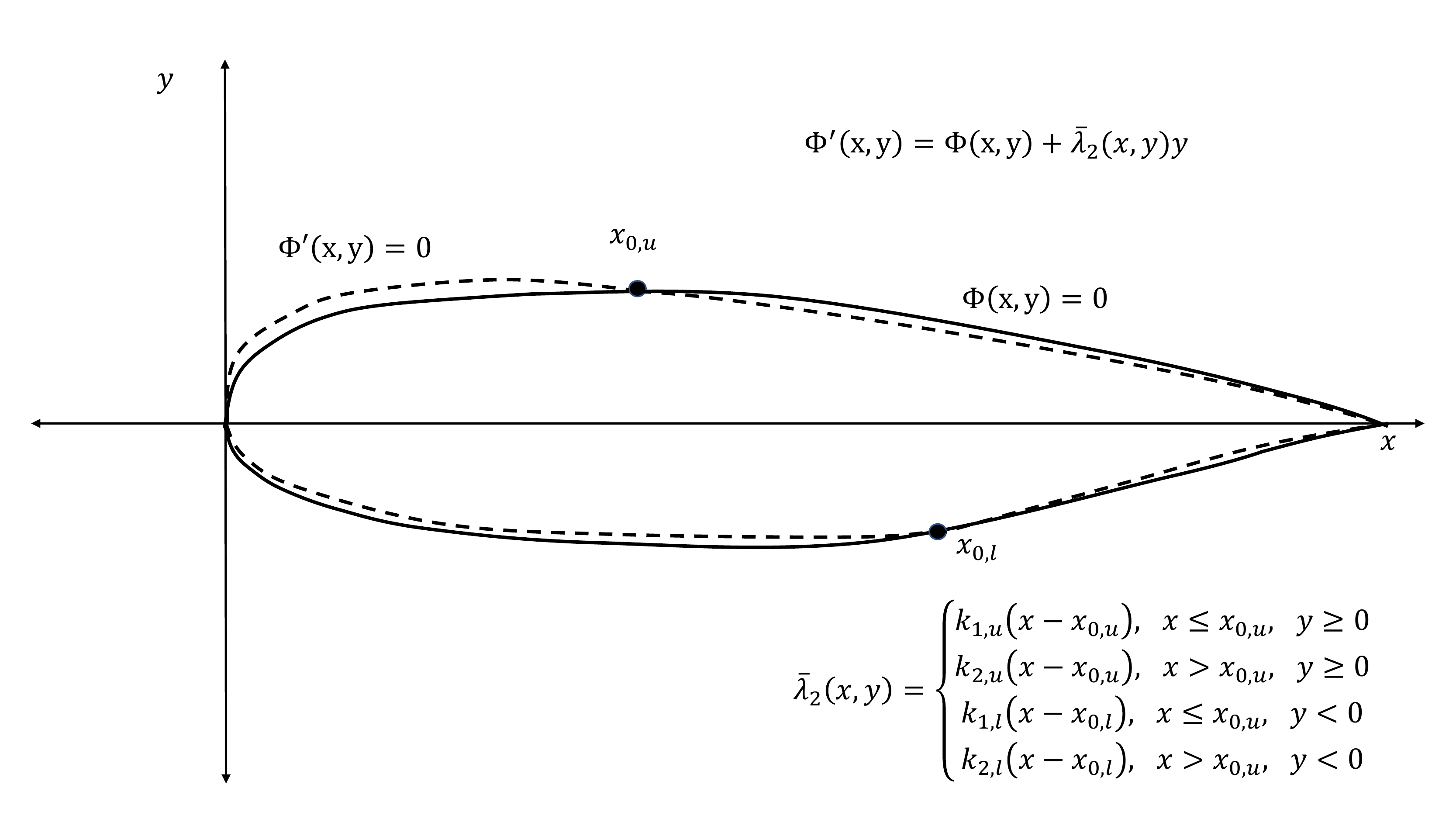}
\caption{Superposition of parameter perturbations to morph a hydrofoil shape}
\label{fig:param_foil_shape_morph}
\end{figure}

Here we will apply the parametric level set approach to morph hydrofoil configurations. Thus, we restrict ourselves to a two-dimensional domain. We also need to consider two other application-specific criteria. First, hydrofoil shapes are inextensible along the $x$-axis during morphing to maintain constant chord length. Thus, we will only consider perturbations along the $y$-axis. Second, we desire the vertical perturbation of the shape to vary along the upper and lower surface of the hydrofoil and also along the chord length. In other words $\Bar{\lambda}_2 \coloneqq \bar{\lambda}_2(x,y)$ instead of a constant value. This is achieved by considering a morphing operation analogous to the vertical displacements on an elastic string about an initial configuration, with zero-displacement boundary conditions at prescribed locations. This is shown in Fig. \ref{fig:param_foil_shape_morph}, where parameters $k_{1,u}$, $k_{2,u}$, $k_{1,l}$, $k_{2,l}$ determine the amplitude of the displacement, and $x_{0,u}$ and $x_{0,l}$ determine the shape of the displacement pattern.

Mathematically, $\Bar{\lambda}_2(x,y)$ can be represented as
\begin{equation}
    \label{eq:param_foil_perturb}
    \Bar{\lambda}_2(x,y) = 
    \begin{cases}
        k_{1,u} \left(x - x_{0,u}\right), &\quad x \leq x_{0,u}, \quad y \geq 0,\\
        k_{2,u} \left(x - x_{0,u}\right), &\quad x > x_{0,u}, \quad y \geq 0,\\
        k_{1,l} \left(x - x_{0,l}\right), &\quad x \leq x_{0,l}, \quad y < 0,\\
        k_{2,l} \left(x - x_{0,l}\right), &\quad x > x_{0,l}, \quad y < 0.
    \end{cases}
\end{equation}
Thus, we propose to morph the hydrofoil smoothly with only six parameters, $k_{1,u}$, $k_{2,u}$, $k_{1,l}$, $k_{2,l}$, $x_{0,u}$ and $x_{0,l}$. A recent article \cite{junior2022intelligent} employing the FFD-based approach required 20 FFD control points for parameterizing the airfoil shape. In contrast, our proposed level set-based shape representation approach requires three to four times lower design variables than the FFD-based approach. Furthermore, the present approach implementation of parametric level sets requires several orders lower parameters than the conventional parametric level set (requiring $(m+1)\times (n+1)$ parameters), especially for a fine level set mesh. Fine level set mesh representation can require anywhere between two to four orders of level set grid points \cite{guirguis2018high,bhatnagar2019prediction} for two-dimensional cases. Thus, we can safely conclude that our present approach requires at least two orders lower design variables than the traditional use of parametric level set representation for topology optimization. The efficiency of parameterizing shapes with a few shape design variables is clearly evident during a gradient-based optimization approach as we require much fewer shape sensitivity computations without resorting to adjoint-based techniques \cite{reuther1996aerodynamic,martins2013review}. For metaheuristic evolutionary optimization techniques lower design variables would significantly reduce the number of design searches required to reach the global optimum \cite{deb2002fast} and could reduce optimization time by orders of magnitude. 

It is important to note that in this study we consider the morphing of a hydrofoil configuration which is at zero degrees angle of attack (see Fig. \ref{fig:param_foil_shape_morph}) for the simplicity of demonstration. However, the proposed perturbation parameters shown in Eq. \ref{eq:param_foil_perturb} can also be applied to hydrofoil configuration at non-zero angles of attack. Such configurations can be easily represented with minor modifications of the perturbation functions to ensure they reach zero at the leading and trailing edges. Non-smooth hydrofoil surfaces containing flaps can also be morphed by the present perturbation approach at the expense of a few additional control points and design variables. However, the number of design variables used would still not increase beyond FFD-based representation for such configurations. The only limitation of the present shape perturbation approach would be when shapes with numerous edges or discontinuous points are considered (e.g., a polygon). For such shapes, the number of morphing design variables required will begin to increase beyond FFD-based techniques with an increase in the number of sharp edges. However, we do not expect such shapes for fluid flows purely because of their poor drag and lift characteristics. Thus, the perturbation approach presented here can be applied for hydrofoil morphing without any loss in generality.  

\subsection{Retrieving morphed boundary from perturbed level sets}
The approach proposed so far can compute the morphed level set functions on a uniform Cartesian grid. However, in order to compute the flow field around the morphed configurations via a CFD solver we need to accurately locate the actual boundary. Usually, the level set mesh is coarser than the discretization required for CFD simulations. Thus, the level set mesh is first refined until $\Delta x^t,\Delta y^t \approx h$, where $\Delta x^t,\Delta y^t$ are the mesh size of the refined Cartesian grid and $h$ is the mesh size at the boundary required for accurate CFD simulation. Subsequently, we compute the level set on the refined Cartesian grid $\bm{\mathrm{X}}^t=\left\{\bm{\mathrm{x}}_1^t, \bm{\mathrm{x}}_2^t ,\ldots, \bm{\mathrm{x}}_{N_t}^t\right\}$ via interpolation of the signed distance functions computed on the control Cartesian grid $\bm{\mathrm{X}}^c$ to $\bm{\mathrm{X}}^t$. This is achieved via the smooth interpolating properties of the RBFs as follows. The signed distances $f_i^t, 1\leq i\leq N_t $, at the refined mesh grid points $\bm{\mathrm{X}}^t$, can be obtained from
\begin{equation}
    \label{eq:refined_sdf}
    \bm{f}^t = \bm{H} \bm{\Lambda},
\end{equation}
where 
\begin{align}
    \label{eq:refined_sdf_eq_coeffs}
    \bm{H} &= \left[\bm{P}' \; \bm{A}'\right]\\
    \bm{f}^t &= {\left[0\; 0 \; 0\; f_1^t \; \ldots \; f_{N_t}^t\right]}^T,\\
    \bm{P}' &=
    \begin{bmatrix}
    1 & x_1^t & \left(1 + \frac{\Bar{\lambda}(x_1^t,y_1^t)}{\lambda_2}\right)y_1^t \\
    \vdots & \vdots & \vdots \\
    1 & x_{N_t}^t & \left(1 + \frac{\Bar{\lambda}(x_{N_t}^t,y_{N_t}^t)}{\lambda_2}\right)y_{N_t}^t
    \end{bmatrix},\\
    {A'}_{i,j} &= \phi \left(\lVert \bm{\mathrm{x}}_j^c - \bm{\mathrm{x}}_i^t \rVert\right), \; 1\leq i \leq N_t, \; 1\leq j \leq N_c.
\end{align}

Since the level set function represents the signed distance from the boundary, we can inspect changes in the sign of the distance function to detect the boundary of the morphed shape. This is achieved by detecting the $y^t$ associated with a change in the sign of the level set at each $x^t$, and subsequently interpolating them to obtain the $\bm{\mathrm{x}}^t_* = (x^t_*,y^t_*)$ which satisfies $\Phi' \left(\bm{\mathrm{x}}^t_* \right)=0$. Assuming $\bar{m}$ and $\bar{n}$ are the spatial discretizations along the $x$ and $y$-axis respectively, of the refined Cartesian mesh, algorithm \ref{alg:interp_to_bndry} shows the steps required to obtain the coordinates of the morphed boundary for each perturbation parameter $\Bar{\lambda}(x,y)$.  
\begin{algorithm}
\caption{Morphed boundary for each perturbed parameter set}\label{alg:interp_to_bndry}
\hspace*{\algorithmicindent} \textbf{Input}: $\Bar{\lambda}(x,y)$ \\
\hspace*{\algorithmicindent} \textbf{Output}: $(x^t_*,y^t_*) $ satisfying $\Phi'(x^t_*,y^t_*)=0$
\begin{algorithmic}[1]
\Procedure{ }{}
\State Obtain refined mesh $\bm{\mathrm{X}}^t$ such that $\Delta x^t,\Delta y^t \approx h$
\State From $\Bar{\lambda}(x,y)$ compute $\bm{H}$ on $\bm{\mathrm{X}}^t$ via Eq. \ref{eq:refined_sdf_eq_coeffs}
\State Compute signed distances $\bm{f}^t$ on $\bm{\mathrm{X}}^t$ via Eq. \ref{eq:refined_sdf}
\For{\texttt{$i \leq \bar{n}$}}
    \For{\texttt{$j \leq \bar{m}-1$}}
    \State $k \gets 0$
    \If{$\sgn\left(\Phi'(x^t_i,y^t_j)\right) \neq \sgn(\Phi'\left(x^t_i,y^t_{j+1})\right)$}
    \State store $\left(\Phi'(x^t_i,y^t_j),\Phi'(x^t_i,y^t_{j+1})\right)$ and $(y^t_j,y^t_{j+})$
    \State $k \gets k+1$
    \EndIf
    \EndFor
    \If{$k>0$}
    \For{\texttt{$l\leq k$}}
    \State Obtain linearly interpolated $(x^t_{i*},y^t_{j*})$
    \EndFor
    \EndIf
\EndFor
\EndProcedure
\end{algorithmic}
\end{algorithm}

\section{Surrogate modelling for fluid flow prediction}
\subsection{Full-order flow field solution}
Here we are interested in predicting the flow-field $\bm{\mathrm{u}}$ around various morphed hydrofoil configurations for various flow conditions. The morphed hydrofoil shape $\partial \Omega'$ is obtained on a refined discretization $\bm{\mathrm{X}}^t_* \in \partial \Omega'$ according to algorithm \ref{alg:interp_to_bndry}, such that $\Phi \left({\bm{\mathrm{X}}_*}^t;\xi \right)=0$. $\xi \in \mathbb{R}^P$ is a set parameters representing the various flow conditions. Once we obtain the explicit hydrofoil shape we can employ our full-order model $\mathcal{F}$ to obtain the flow-field $\bm{\mathrm{U}}$ on the CFD mesh
\begin{equation}
    \label{eq:FOMsoln}
    \bm{\mathrm{U}} \left(\bm{\mathrm{X}};\xi \right) = \mathcal{F}\left(\bm{\mathrm{X}},\bm{\mathrm{X}}_*^t;\xi \right).
\end{equation}
The flow field $\bm{\mathrm{U}}$ is interpolated on the control level set grid $\bm{\mathrm{X}}^c$ to obtain $\bm{\mathrm{U}}^c= \left\{\bm{\mathrm{u}}_1^c,\bm{\mathrm{u}}_2^c,\ldots,\bm{\mathrm{u}}_{N_c}^c\right\}$. Thus, the interpolated flow field only depends implicitly on $\Phi'\left(\bm{\mathrm{X}}^c\right)$. For most complex shapes and flow conditions, closed-form solutions to the flow field are not available and $\mathcal{F}$ is obtained either via computational solvers or experiments. 


\subsection{Data-driven surrogate modelling}
The objective of the data-driven surrogate modelling is to obtain an approximate map $\mathcal{G}$ of the full-order model $\mathcal{F}$. $\mathcal{G}$ is learned only from a set of known input data $\Phi' \left(\bm{\mathrm{X}}^c \right)$ and observed output data $\bm{\mathrm{U}}^c \left(\bm{\mathrm{X}}^c;\Phi' \left(\bm{\mathrm{X}}^c \right), \xi \right)$ via a set of real-valued learnable parameters $\Theta$. Once we have learned $\mathcal{G}$ it can be used for predicting the flow field for out-of-training parametric level sets $\tilde{\Phi'}$,
\begin{equation}
    \label{eq:datadrivenpred}
    \tilde{\bm{\mathrm{U}}^c} \left(\bm{\mathrm{X}}^c;\xi \right) = \mathcal{G}\left( \tilde{\Phi'} \left(\bm{\mathrm{X}}^c \right);\Theta \right).
\end{equation}
Being data-driven, the surrogate map $\mathcal{G}$ is agnostic to how the training data is obtained. Hence, it can be employed for both numerically or experimentally observed data or a combination of both.

\subsection{Convolutional encoder-decoder network}
The data-driven surrogate model is obtained here via the convolutional encoder-decoder, a deep learning architecture developed completely via CNNs. The model consists of a CNN encoder and a CNN decoder. The CNN encoder generates a dimensionality-reduction map from the high-dimensional input data on a Euclidean domain to low-dimensional latent variables. The CNN decoder subsequently provides a dimensionality-expansion map from the low-dimensional latent variables to the high-dimensional output, which also lies on the same Euclidean domain as the input. Thus, the convolutional encoder-decoder is the data-driven operator $\mathcal{G}$ in Eq. \ref{eq:datadrivenpred}, operating on high-dimensional implicit shape representation $\Phi'\in \mathbb{R}^{N_c}$ and observed variable of interest $\bm{\mathrm{U}}^c \in \mathbb{R}^{N_c}$. Mathematically we can represent the convolutional encoder-decoder $\mathcal{G}$ as the composition of the CNN encoder $\mathcal{E}$ and decoder $\mathcal{D}$,
\begin{equation}
    \label{eq:data-driven_learning}
    \begin{split}
        \bm{\mathrm{A}}^c\left(\xi \right) =& \mathcal{E}\left(\Phi'\left(\bm{\mathrm{X}}^c \right);\xi,\bm{\theta}_\mathcal{E} \right),\\
        \bm{\mathrm{U}}^c\left(\bm{\mathrm{X}}^c;\xi \right) =&\mathcal{D}\left(\bm{\mathrm{A}}^c\left(\xi \right);\bm{\theta}_\mathcal{D}\right)\\
        \mathcal{G} =& \left(\mathcal{E}\circ\mathcal{D};\bm{\mathrm{\Theta}}\right).
    \end{split}
\end{equation}
Here $\bm{\mathrm{A}}^c \in \mathbb{R}^P$ and $\bm{\mathrm{\Theta}} = \left(\bm{\theta}_\mathcal{E},\bm{\theta}_\mathcal{D}\right)$. The dimensional compression performed by the CNN encoder via its deep hierarchical structure usually results in a much lower number of required trainable parameters as usually, $P\ll N_c$. 

\begin{figure}[ht]
\centering
\includegraphics[width=0.96\textwidth]{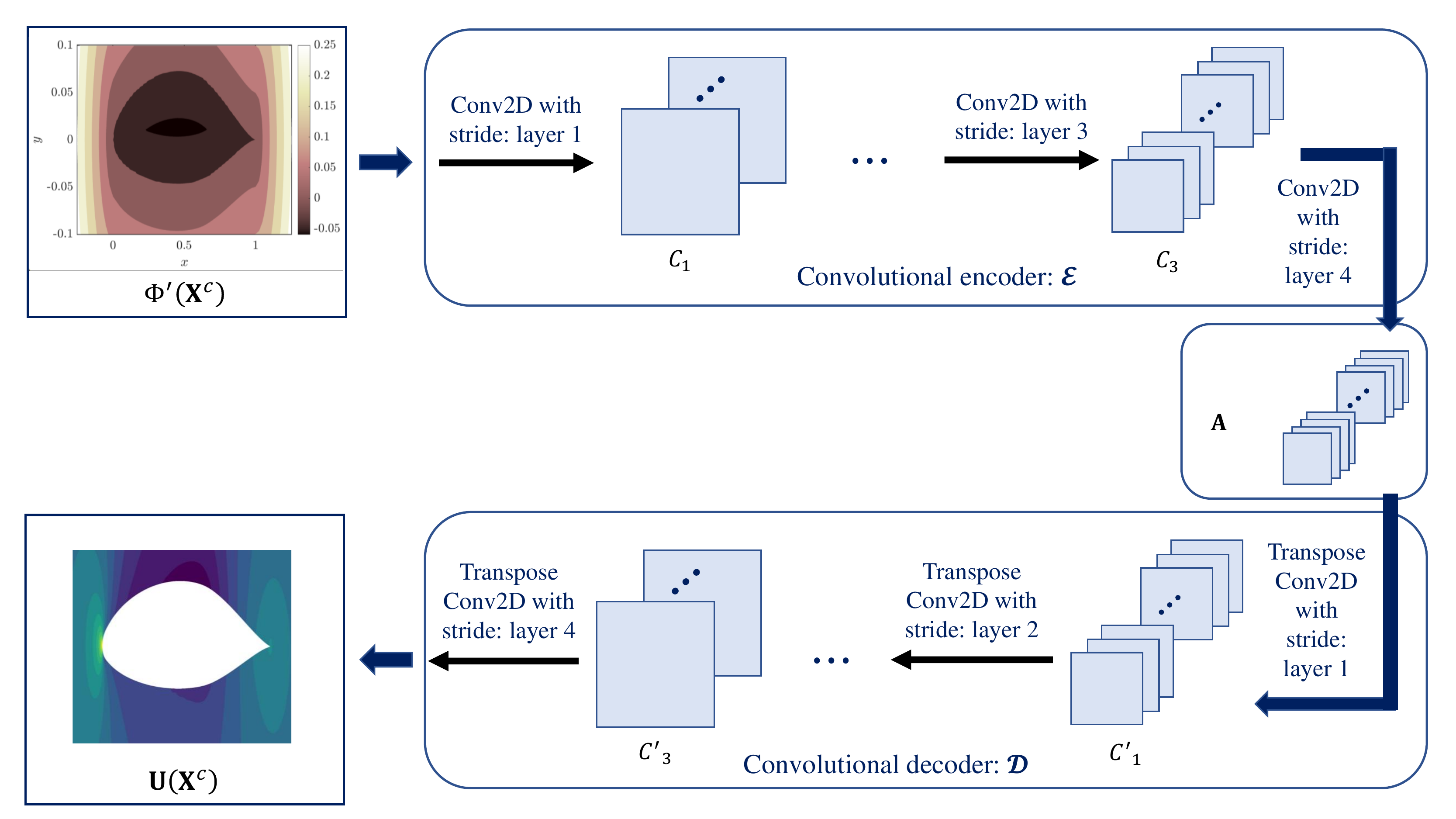}
\caption{Framework of the convolutional encoder-decoder network}
\label{fig:CNN_enc_dec}
\end{figure}

A schematic showing the CNN encoder-decoder architecture is shown in Fig. \ref{fig:CNN_enc_dec}. The convolutional encoder is a composition of four convolutional layers, which progressively reduces the spatial dimension of the input. However, the reduced dimensional feature spaces obtained from each convolutional encoder layer ($C_1, C_2$, etc.) have an increasing number of convolutional filters. The final output of the convolutional encoder, $\bm{\mathrm{A}}$, is subsequently passed to the convolutional decoder, which is a composition of four transposed convolutional layers. These layers operate exactly opposite to the convolutional encoder layers, resulting in progressively expanding spatial dimension and reduction of convolutional filters. The final output obtained from the decoder has the same dimensionality as the input.

Each convolutional layer performs a convolutional operation with strides and nonlinear activation. The convolution operation provides translation equivariance to the convolutional encoder \cite{bronstein2017geometric}. The strides in the convolution behave similarly to a pooling operation \cite{bronstein2017geometric} and provide translation invariance. Thus, the convolutional encoder is not just a regression model as its architecture provides geometric priors to the mapping. In an analogous manner, the transpose convolution and upsampling operations in each transposed convolutional layer of the decoder also preserve the equivariance and invariance of the encoder-decoder map to translations. A discussion on how such geometric priors improve generalization for learning convection-dominated fluid dynamics and their physical interpretability is provided in Ref. \cite{mallik2022assessment}. It is worth mentioning that the encoder-decoder architecture selected here is comprised completely of convolutional layers to fully preserve the geometric priors of the convolutional encoder and decoder. Other convolutional encoder-decoder-based architectures have also recently been developed \cite{bhatnagar2019prediction} for flow prediction, where the feature spaces from convolutional layers have been flattened via fully connected layers into a low-dimensional set of latent variables. However, it has been explained \cite{bronstein2017geometric} that such flattening leads to a loss of the geometric priors like translation equivariance and invariance of the network. 

\subsection{Training surrogate model}
The convolutional encoder-decoder surrogate model has to be trained offline before it can be employed online for shape optimization. Let $\bm{b}=\left\{b_1,b_2,\ldots,b_6\right\}, \bm{b} \in \mathbb{R}$, be the set of six parameters defined earlier, which will be used to perturb the baseline hydrofoil shape. We define $\bm{b}_u$ and $\bm{b}_l$ as the set of upper and lower bounds, respectively, such that $\bm{b}_l \leq \bm{b} \leq \bm{b}_u$. We subsequently select $N$ sampling parameter set $\bm{b}_i, i=1, \ldots, N$, within the defined bounds. Algorithm \ref{alg:inp-out_data_conv-enc-dec} explains how to obtain the training input data set ${\bm{\Phi}'}^N=\left\{{\Phi'}_1,{\Phi'}_2,\ldots,{\Phi'}_N\right\}$ and output data set $\bm{\mathrm{U}}^N= \left\{\bm{\mathrm{U}}^c_1,\bm{\mathrm{U}}^c_2,\ldots,\bm{\mathrm{U}}^c_N\right\}$. The convolutional encoder-decoder model is then trained via standard neural network training techniques.

\begin{algorithm}
\caption{Offline convolutional encoder-decoder training}\label{alg:inp-out_data_conv-enc-dec}
\hspace*{\algorithmicindent} \textbf{Input}: $\bm{b}_l$, $\bm{b}_u$ \\
\hspace*{\algorithmicindent} \textbf{Output}: ${\bm{\Phi}'}^N$, $\bm{\mathrm{U}}^N$
\begin{algorithmic}[1]
\Procedure{ }{}
\State Obtain $N$ sampled sets: $\bm{b}_l \le \bm{b}_i \le \bm{b}_u, \ i=1, \ldots, N$
\For{\texttt{$i \leq N$}}
    \State For each $\bm{b}_i$ obtain $\Bar{\lambda}_{2,i}(x,y)$ via Eq. (\ref{eq:param_foil_perturb})
    \State Using $\Bar{\lambda}_{2,i}(x,y)$ and baseline shape, obtain ${\Phi'}_i$ via Eqs. (\ref{eq:perturb_approach}), (\ref{eq:add_constrnts}-\ref{eq:Wendland's_2D_basis})
    \State Obtain $(x^t_*,y^t_*)$ for ${\Phi'}_i$ via Algorithm \ref{alg:interp_to_bndry}
    \State Obtain CFD solution for $(x^t_*,y^t_*)$ 
    \State Interpolate CFD solution on $\bm{\mathrm{X}}^c$ to obtain $\bm{\mathrm{U}}^c_i$ 
\EndFor
\EndProcedure
\end{algorithmic}
\end{algorithm}

\section{Shape optimization framework}
The shape optimization framework consists of two phases. One is the optimization phase, which involves formulating the optimization problem and performing the gradient-based optimization. The second part involves the online application of the convolutional encoder-decoder model in the optimization.

\subsection{Non-intrusive shape optimization}
The shape optimization involves the minimization of the objective function $\mathcal{J}$, subject to a set of inequality constraints $\mathcal{C_I}$. Both the objective and the constraints are a function of the flow field $\bm{\mathrm{U}}$ and the shape perturbation parameter set $\bm{b}$. This can be mathematically formulated as,
\begin{equation}
    \begin{split}
        \label{eq:gen_opt_formulation} 
        \underset{\bm{b} \in \mathscr{V}}{\text{minimize}} & \quad \mathcal{J} \biggl(\bm{\mathrm{U}}\Bigl(\mathrm{\Phi'}\bigl(\bm{\mathrm{X}};\bm{\mathrm{X}}^*(\bm{b})\bigr)\Bigr), \bm{\mathrm{X}}^*(\bm{b})\biggr)\\
        \text{subject to} & \quad
        \mathcal{C_I} \left(\bm{\mathrm{U}}\left(\mathrm{\Phi'}\left(\bm{\mathrm{X}};\bm{\mathrm{X}}^*\left(\bm{b}\right)\right)\right), \bm{\mathrm{X}}^*\left(\bm{b}\right)\right)\leq 0,
    \end{split}
\end{equation}
where $\Phi\bigl(\bm{\mathrm{X}}^*(\bm{b})\bigr)=0$. Since $\bm{b}$ is a set of six design variables, $\mathscr{V}$ represents a six-dimensional parameter space of all possible combinations of design variables. Eligible solutions to this problem are the ones that satisfy all constraints and are members of the set $\bm{\Gamma}$
\begin{equation}
    \label{eq:opt_bounds}
    \bm{\Gamma}=\{\wp \left({\left[b_1,\ldots,b_6\right]}^T\right)|\bm{b_l}\leq\bm{b}\leq\bm{b_u}\},
\end{equation}
which contains all variable permutations $\wp$ of $\bm{b}$ between their lower and upper bounds, $\bm{b_l}$ and $\bm{b_u}$, respectively. 

The shape optimization methodology presented above is a generic formulation applicable to any non-intrusive objective function and constraint evaluation approach. Thus, it allows the computation of objectives and constraints via both a full-order model and a surrogate model. Furthermore, the non-intrusive approach adopted here makes the optimization framework compatible with any generic optimization algorithm. 

In the present study, we employ a gradient-based optimization algorithm. The gradient of the objective function with respect to the shape morphing variables and the shape sensitivity of the constraint
is computed directly via any routine numerical differentiation scheme. Such computation of sensitivities is termed as monolithic differentiation in the gradient-based shape optimization literature \cite{martins2013review} and is employed when the computational model is treated as a ``black-box" model.

\subsection{Online application of convolutional encoder-decoder}
The trained convolutional encoder-decoder $\mathcal{G}$ is employed online during each iteration of the shape optimization for the computation of the objective function and the constraints. At each iteration, we know our level set representation $\Phi' \left(\bm{\mathrm{X}} \right)$ of the perturbed hydrofoil boundary. With the trained parameters $\Theta$, we can then predict our flow field via Eq. \ref{eq:datadrivenpred}. 
The whole shape optimization process, including the application of the convolutional encoder-decoder surrogate model for objective and constraint evaluation, is shown in algorithm \ref{alg:shape_opt_conv-enc-dec}.
\begin{algorithm}
\caption{Shape optimization with convolutional encoder-decoder model}\label{alg:shape_opt_conv-enc-dec}
\hspace*{\algorithmicindent} \textbf{Input}: $\Phi\left(\bm{\mathrm{X}}\right)$, $\bm{b}_l$, $\bm{b}_u$ \\
\hspace*{\algorithmicindent} \textbf{Output}: Optimal shape perturbation parameter set $\bm{b}^* $
\begin{algorithmic}[1]
\Procedure{ }{}
\State Initiate optimization with $\bm{b} \subset \bm{\mathrm{\Gamma}} $
\For{\texttt{$i \leq N_{\max} $}}
    \State Compute $\Bar{\lambda}_2(\bm{\mathrm{X}})$ with $\bm{b}$ via Eq. \ref{eq:param_foil_perturb}
    \State Compute $\Phi'\left(\bm{\mathrm{X}}\right)$ from $\Phi\left(\bm{\mathrm{X}}\right)$ and $\Bar{\lambda}_2(\bm{\mathrm{X}})$ via Eq. \ref{eq:perturb_approach} 
    \State Compute $\bm{\mathrm{U}} \left(\bm{\mathrm{X^c}}\right)$ via trained network $\mathcal{G}$ (Eq. \ref{eq:datadrivenpred})
    \State Compute $\mathcal{J} \left(\bm{\mathrm{U}}\left(\mathrm{\Phi'}\left(\bm{b}\right)\right), \bm{b}\right)$ and $\mathcal{C_I} \left(\bm{\mathrm{U}}\left(\mathrm{\Phi'}\left(\bm{b}\right)\right), \bm{b}\right)$
    \If{convergence criteria met}
        \State $\bm{b}^*=\bm{b}$
    \Else{}
        \State Continue optimization
    \EndIf
\EndFor
\EndProcedure
\end{algorithmic}
\end{algorithm}

\begin{figure}[ht]
\centering
\includegraphics[width=0.96\textwidth]{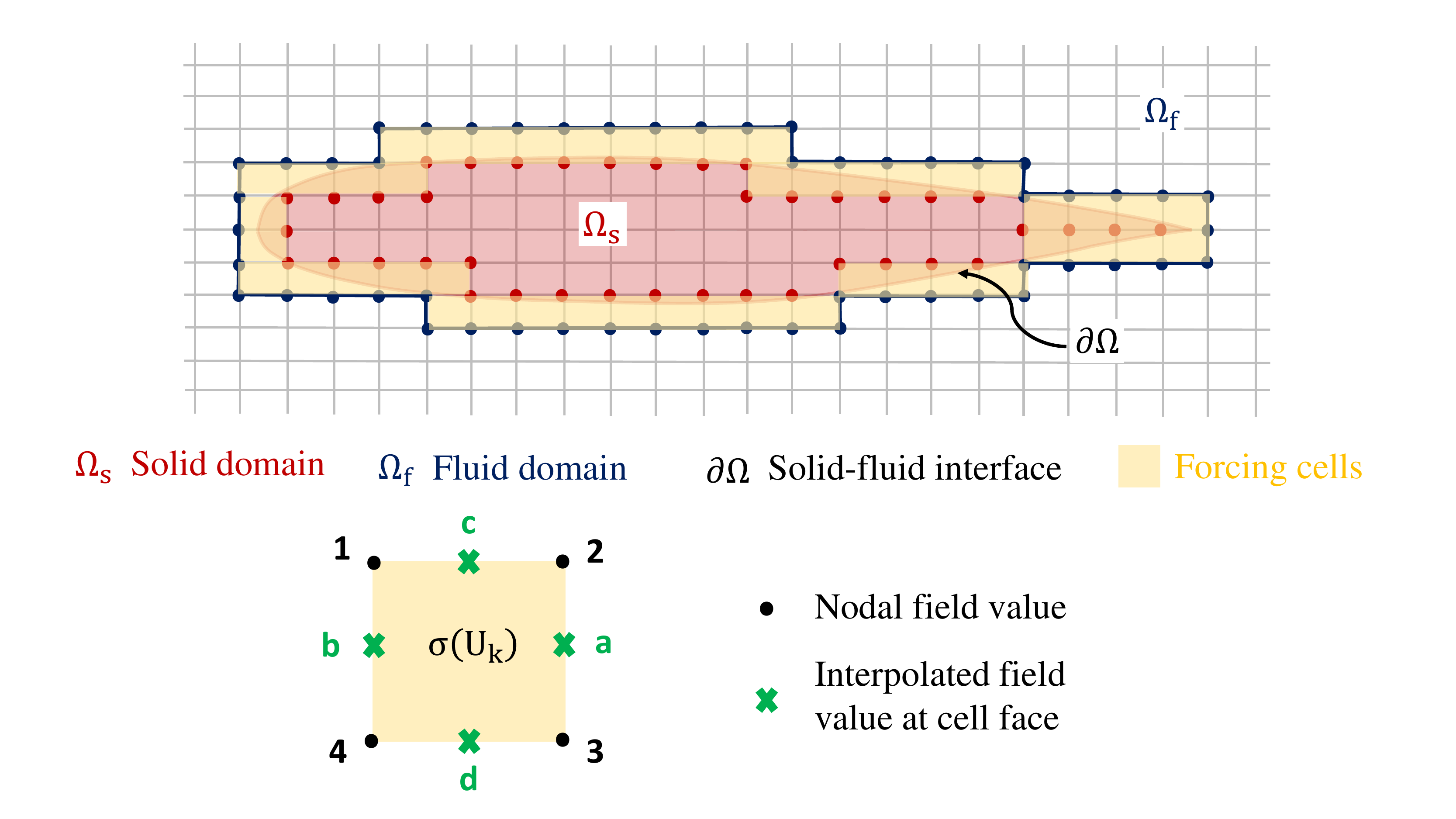}
\caption{Identification of cells at the hydrofoil-fluid interface where field data is integrated to obtain force components}
\label{fig:identify_force_cells}
\end{figure}

The objective and constraint functions often involved in shape optimization of aero/hydrofoils are the lift and drag coefficients. We need to compute them from the flow fields provided by the convolutional encoder-decoder model on the uniform Cartesian level set control mesh $\bm{\mathrm{X}}^c$. However, the boundary of the hydrofoil will not always lie on a grid point of $\bm{\mathrm{X}}^c$, which will pose a challenge to integrate the force at the interface from field values on the level set mesh. To address this problem of force calculation on the hydrfoil interface, we first identify all the interface cells for a given solid boundary and sum the force experienced by these cells. Fig. \ref{fig:identify_force_cells} shows how such forcing cells on the solid-fluid interface are identified. The force computation process closely follows the one employed in Ref. \cite{bukka2021assessment} and is explained briefly below.

The force in each interface cell is the product of the Cauchy stress tensor $\upsigma\left(\bm{\mathrm{U}}_k\right)$ and the area of the cell at a specific location $k$. $\bm{\mathrm{U}}_k = {\left(\mathrm{p}_k, \bm{\mathrm{v}}_k\right)}^T$ comprises of the pressure and velocity field variables at cell $k$. The discrete force $\mathrm{f}_k$ at cell $k$ can thus be evaluated from the stress tensor as,
\begin{equation}
    \label{eq:force_comp_interface_cell}
    \mathrm{f}_k = \left(\mathrm{\upsigma}_{a;k}-\mathrm{\upsigma}_{b;k}\right)\cdot \mathrm{n}_x \Delta y + \left(\mathrm{\upsigma}_{c;k}-\mathrm{\upsigma}_{d;k}\right)\cdot \mathrm{n}_y \Delta x,
\end{equation}
where $\mathrm{n}_x$ and $\mathrm{n}_y$ are the unit vectors in the $x$ and $y$ directions, respectively. $\Delta x$ and $\Delta y$ are the cell dimensions in the $x$ and $y$ directions, respectively. To obtain the discrete force in each cell we need to compute the stress tensor at the mid-face locations of cell $k$ from the nodal locations (as shown in Fig. \ref{fig:identify_force_cells}) via finite difference interpolation shown below
\begin{equation}
    \label{eq:mid-face_val_comp}
    \begin{split}
        \mathrm{\upsigma}_{a;k} = \frac{\mathrm{\upsigma}_{2;k} + \mathrm{\upsigma}_{3;k}}{2}, \quad \mathrm{\upsigma}_{b;k} = \frac{\mathrm{\upsigma}_{4;k} + \mathrm{\upsigma}_{1;k}}{2}, \\
        \mathrm{\upsigma}_{c;k} = \frac{\mathrm{\upsigma}_{1;k} + \mathrm{\upsigma}_{2;k}}{2}, \quad \mathrm{\upsigma}_{d;k} = \frac{\mathrm{\upsigma}_{3;k} + \mathrm{\upsigma}_{4;k}}{2}.
    \end{split}
\end{equation}
The Cauchy stress tensor can be expanded for any point $*$ inside the cell $k$ as $\upsigma_{*;k} = -\mathrm{p}_{*;k} \bm{\mathrm{I}} + 2 \mu \bm{\mathrm{E}}_{*;k}$, where $\mathrm{p}_{*;k}$, $\bm{\mathrm{I}}$, $\mu$ and $\bm{\mathrm{E}}_{*;k}$ are the fluid pressure, identity tensor, fluid viscosity and fluid strain rate tensor, respectively. In the present case, we only consider the pressure component of the stress tensor and neglect the viscosity component similar to Ref. \cite{bukka2021assessment}. With this assumption, and using Eqs. \ref{eq:force_comp_interface_cell} and \ref{eq:mid-face_val_comp}, the expression for the discrete force at each cell $k$ becomes
\begin{multline}
    \label{eq:discrete_force}
    \mathrm{f}_k = \begin{bmatrix} 
	\mathrm{p}_{b;k}-\mathrm{p}_{a;k} & 0 \\
	0 & \mathrm{p}_{b;k}-\mathrm{p}_{a;k}
	\end{bmatrix} \cdot 
    \begin{bmatrix}
    1 \\
    0
    \end{bmatrix} \Delta y + \\
    \begin{bmatrix} 
	\mathrm{p}_{d;k}-\mathrm{p}_{c;k} & 0 \\
	0 & \mathrm{p}_{d;k}-\mathrm{p}_{c;k}
	\end{bmatrix} \cdot 
    \begin{bmatrix}
    0 \\
    1
    \end{bmatrix} \Delta x.
\end{multline}
The total force over $N_f$ such cells can be evaluated as
\begin{equation}
    \mathrm{F}_B = \sum_{k=1}^{N_f} \mathrm{f}_k.
\end{equation}

In Ref. \cite{bukka2021assessment}, a correction to the force computed via integration of the pressure in the interface cell was performed. This enables some recovery of the loss in resolution of the field data near the boundary layer due to a coarse level set mesh. However, such a correction strategy cannot be applied here as unlike that article, the hydrofoil boundary here will change continuously during the online shape optimization stage. Thus, any correction obtained during the offline training stage for load recovery cannot be implemented during the online stage. Instead, we have refined the discretization of our level-set grid in the $y$ direction to better resolve the field data variation near the boundary layer. The convergence criteria for such $y$-directional level set mesh refinement and the residual error expected in the force computation will be further discussed in the results section.

\section{Test problem}
In this article, we will present the complete methodology for the convolutional encoder-decoder-based shape optimization of the NACA66 hydrofoil. We will briefly discuss the implementation of the parametric level set shape perturbation approach, the full-order simulations and the convolutional encoder-decoder model training, and the optimization problem. 

\subsection{Training input and output data sets}
The level set of the NACA66 hydrofoil obtained on the level set control mesh was perturbed by a large set of design parameters sampled from the parameter bounds to obtain the training input data set. The perturbed hydrofoil boundaries were obtained from these perturbed level sets via algorithm \ref{alg:interp_to_bndry}. Reynolds-averaged Navier-Stokes simulations with the Spalart Allmaras turbulence model were subsequently performed for these hydrofoil shapes. The flow field data computed on CFD mesh for each of these perturbed hydrofoil shapes were interpolated on the uniform Cartesian level set control mesh via Matlab's linear interpolation scheme. These interpolated flow-field data are the output training set for their perturbed level set counterparts.

\subsection{Full-order solution}
\begin{figure}[ht]
\centering
\includegraphics[width=0.96\textwidth]{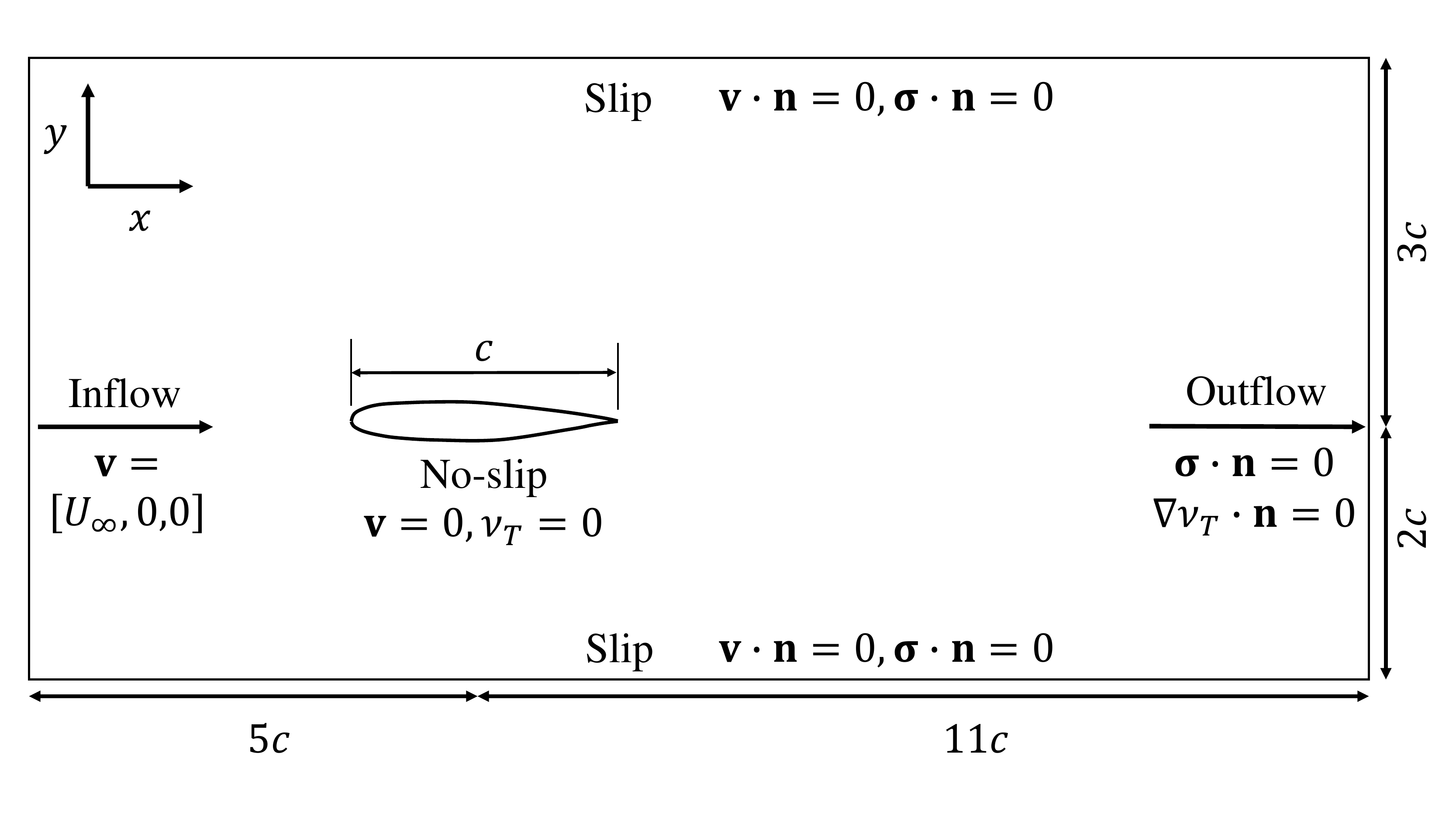}
\caption{Representative computational domain and associated boundary conditions for flow over NACA66 hydrofoil}
\label{fig:comp_domain_with_bc_foil}
\end{figure}

\begin{figure}[ht]
\centering
\includegraphics[width=0.96\textwidth]{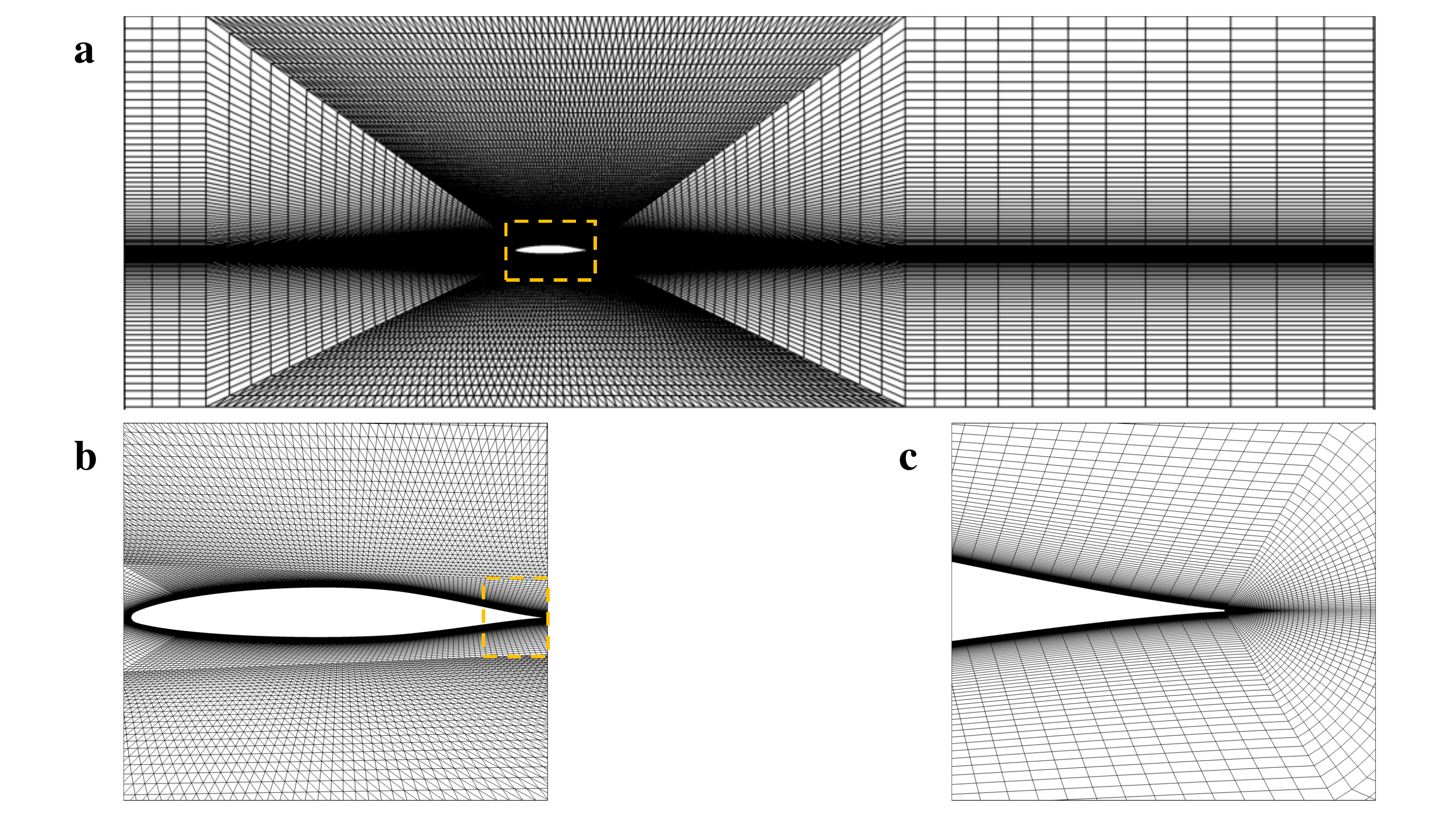}
\caption{Computational mesh used for the flow simulations: (a) Full domain, (b) 16 times magnified - around NACA66 hydrofoil, (c) 100 times magnified - around trailing edge}
\label{fig:comp_mesh}
\end{figure}

We will first discuss how the CFD simulations were performed to obtain the nonlinear flow field data for various hydrofoil configurations obtained via morphing the NACA66. The Reynolds-averaged Navier Stokes simulations were performed on a three-dimensional computational domain whose representative two-dimensional slice is shown in Fig. \ref{fig:comp_domain_with_bc_foil} with the boundary conditions. A Dirichlet velocity boundary condition equal to freestream velocity was applied at the inlet with a natural traction-free outlet condition at the flow exit. A symmetric boundary condition is used on the top and bottom surfaces. Here $\nu_T$ represents the kinematic turbulence viscosity. The hydrofoil has a span of $0.1c$ with periodic boundary conditions on the sides, thus leading to two-dimensional flow behaviour. The computational mesh on the domain is shown in  Fig.\ref{fig:comp_mesh} (a) along with magnifications around the hydrofoil (Fig.\ref{fig:comp_mesh} (b)) and the hydrofoil trailing edge (Fig.\ref{fig:comp_mesh} (c)). The mesh consists of 30,674 hexahedral elements and 16,896 prism elements. A target $y^+=y u_\tau/\nu=1$ 
was maintained in the discretization of the hydrofoil boundary layer, where $y$ is the height of the first node from the wall, $u_\tau$ is the friction velocity and $\nu$ is the kinematic viscosity of the fluid.

As discussed earlier, in this study we will consider the NACA66 hydrofoil at zero degrees angle of attack for simplicity in demonstrating the parametric level set morphing techniques. It is worth mentioning that the NACA66 is not a symmetric hydrofoil and shows non-negligible lift and drag coefficients at zero degrees. Thus, even at zero degrees angle of attack, the morphed shapes can represent the lift and drag characteristics of the hydrofoil at low angles of attack where flow separation is not expected, without any loss in generality. The fluid conditions resembled that of water at $25^\circ C$. The flow Reynolds number was $2.0 \times 10^6$ and $U_\infty=1 m s^{-1}$ was considered. The computations were performed by a finite-element solver employed in previous studies \cite{joshi2017variationally, kashyap2021robust}. The flow field was iteratively converged to its steady-state solution and standard mesh convergence practice was followed to obtain a mesh-independent solution.

\subsection{Convolutional encoder-decoder training}
The convolutional encoder-decoder network was trained over the input and output data sets with the Adam optimization algorithm \cite{kingma2014adam} in Matlab's deep learning toolbox \cite{matlabdl}. Since the flow field solutions are predicted by the convolutional encoder-decoder in a pixelated format on a Cartesian grid, their accuracy will be measured by the Structural Similarity Index Measure (SSIM). SSIM is a statistical measure employed primarily for comparing two images \cite{wang2004image}. However, they have also recently been employed on pixelated data sets to assess CNN prediction accuracy \cite{mallik2022predicting}. An SSIM of 1.0 between two images indicates that the two images are identical, whereas 0 implies no similarity. Thus, we will frequently use SSIM in the results section to compare the predicted solutions to their true counterparts obtained on the same level set mesh.  

\subsection{Shape optimization}
The shape optimization will be  performed with to minimize the drag coefficient $c_d$ with respect to the implicit shape perturbation variables $\bm{b}$. The objective is subject to a constraint on the lift coefficient $c_l$, which must lie within a design range $c_l^* \pm \epsilon$. A second optimization constraint is the hydrofoil thickness to chord ratio $t_c$, which must be greater than a prescribed thickness ratio $t_c^*$. Thus, following the general optimization problem formulation shown in Eq. \ref{eq:gen_opt_formulation}, we have for the present case
\begin{equation}
    \begin{split}
        \label{eq:case_opt} 
        \mathcal{J} =&  c_d \biggl(\bm{\mathrm{U}}\Bigl(\mathrm{\Phi'}\bigl(\bm{\mathrm{X}};\bm{\mathrm{X}}^*(\bm{b})\bigr)\Bigr), \bm{\mathrm{X}}^*(\bm{b})\biggr)\\        \mathcal{C_I}(1) =& \bigg| c_l^* - c_l \biggl(\bm{\mathrm{U}}\Bigl(\mathrm{\Phi'}\bigl(\bm{\mathrm{X}};\bm{\mathrm{X}}^*(\bm{b})\bigr)\Bigr), \bm{\mathrm{X}}^*(\bm{b})\biggr) \bigg| - \epsilon\\
        \mathcal{C_I}(2) =& t_c^* - t_c\bigl(\bm{\mathrm{X}}^*(\bm{b})\bigr).
    \end{split}
\end{equation}

\section{Results}
In this section, we will present the convolutional encoder-decoder training and validation results. The shape optimization results with the trained network will also be presented and discussed. However, first we will demonstrate how the hydrodynamic forces were retrieved from the field variables on the level set mesh, and how the retrieved forces were employed to select the level set mesh.

\subsection{Force retrieval and level set mesh selection}
The lift and drag forces were retrieved from the interpolated pressure field as explained earlier in the article. Since we are not considering any correction for the loss in resolution \cite{bukka2021assessment} during the force retrieval, the accuracy of the lift and drag computation will depend on the level set control mesh discretization employed for interpolating the full-order pressure field. Various mesh sizes were selected, and the lift and drag were computed from the level set pressure field for ten representative set samples via the force retrieval method presented earlier. These were compared to the force coefficients directly obtained from the CFD solver for these ten representative designs. 

\begin{figure}[ht]
\centering
\includegraphics[width=0.78\textwidth]{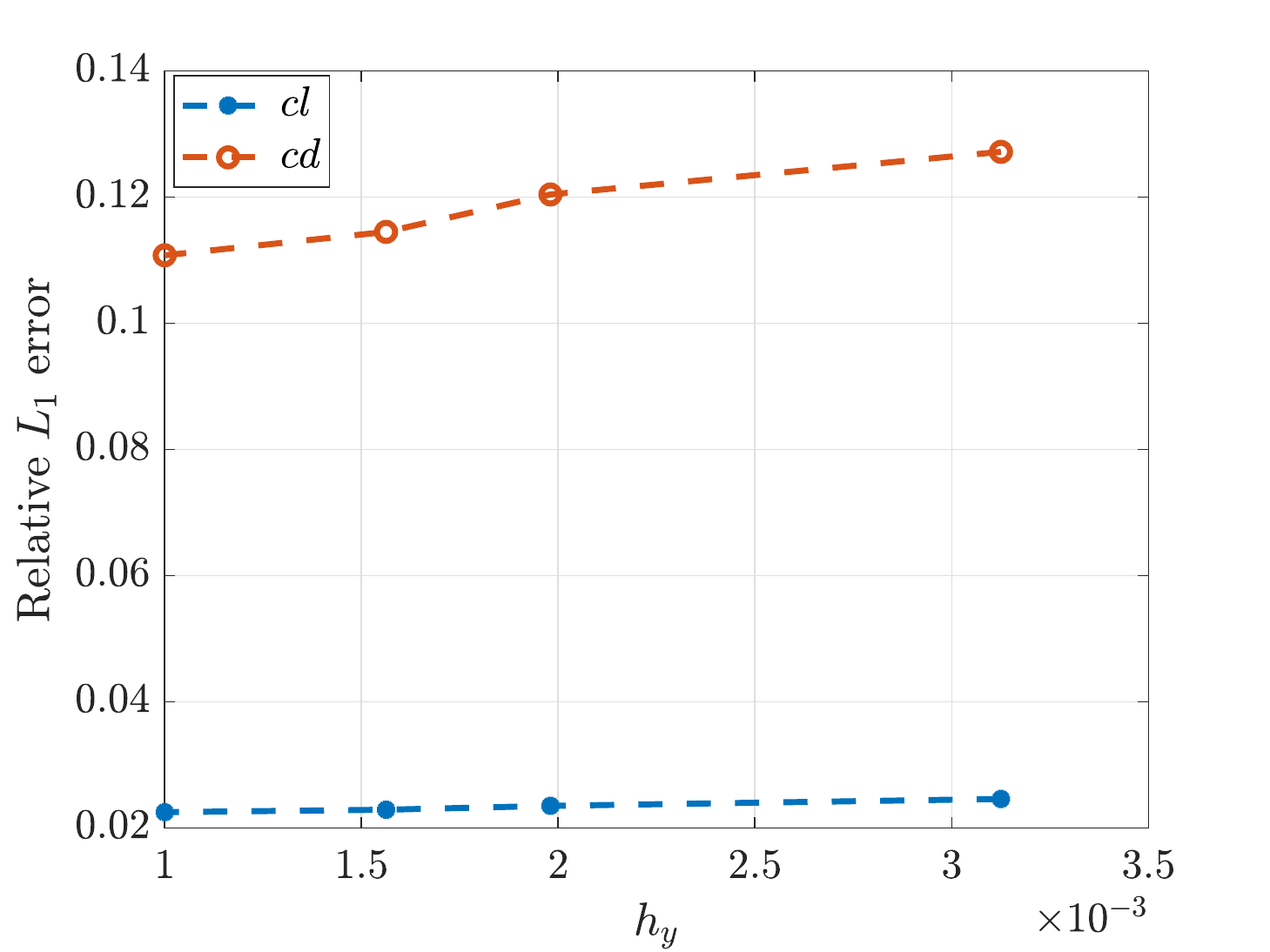}
\caption{$L_1$ error of the retrieved lift and drag coefficients for various mesh sizes}
\label{fig:l1_retrieved_forcecoeffs}
\end{figure}

The relative mean $L_1$ error in both the lift and drag retrieved over the ten representative samples compared to the CFD-computed results are shown in Fig. \ref{fig:l1_retrieved_forcecoeffs} for various mesh sizes $h_y$. $h_y$ represents the grid dimension along the $y$-axis for mesh dimensions $61\times 65$, $61\times 101$, $61\times 129$ and $61\times 201$. As we can see, the lift computed from the interpolated pressure field is quite accurate even with the most coarse interpolation mesh. However, the drag prediction error is almost an order higher than the lift prediction error. The error in drag prediction reduces very slowly as we refine the mesh. Thus, we will consider the $61\times 129$ mesh as adequate as further mesh refinement only increases the computational cost without significant improvement in the drag prediction. It is important to note that the refinement was performed along the $y$-axis to better resolve the boundary layer on the interpolated mesh. Future studies can investigate if including the viscous effects in the Cauchy stress tensor would provide any improvement in the force retrieval.

\subsection{Convolutional encoder-decoder training and validation}
We now discuss the convolutional encoder-decoder training and validation. The convolutional encoder-decoder network was trained with 453 sets of perturbed level-set shape representations as input and their corresponding pressure field interpolated on the level-set grid as output. Another 64 input-output data sets were considered for validation. The objective of the network training is to minimize the mean square errors between the training predictions and the target. However, we must also consider the prediction errors on the validation set to ensure adequate network generalization. Thus, we will simultaneously inspect the network training and validation, which relies on several network hyperparameters. We considered several hyperparameters like network learning rate, network size and the number of training epochs. Here we only show the effect of the number of training epochs on the network training and validation accuracy as it proved to be the most important hyperparameter. 

\begin{figure}[ht]
\centering
\includegraphics[width=0.96\textwidth]{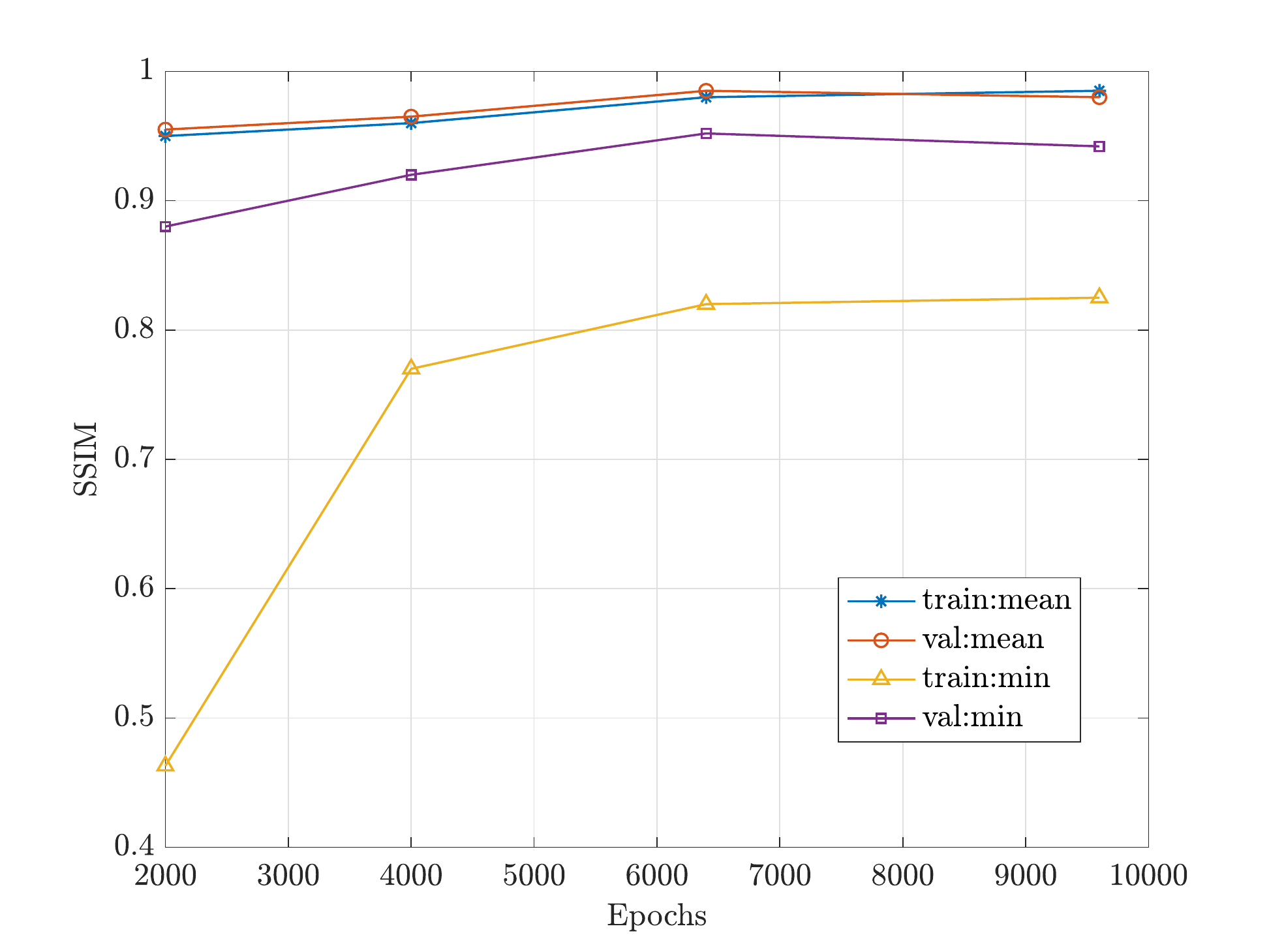}
\caption{SSIM of convolutional encoder-decoder training and validation predictions with respect to the number of training epochs}
\label{fig:SSIM_CNN-enc_dec_train-val}
\end{figure}

Following the application of SSIM in Ref. \cite{mallik2022predicting} for comparing CNN prediction accuracy, here we present the training and prediction accuracy of the convolutional encoder-decoder with respect to the number of training epochs in Fig. \ref{fig:SSIM_CNN-enc_dec_train-val}. Both the mean SSIM over the training and validation sets and the minimum SSIM were considered. We can see that while the mean and minimum SSIM of the training predictions improved slightly on training the network beyond 6400 epochs, the validation SSIM began to decrease. This indicates overfitting of the network on training it beyond 6400 epochs. It is also noted that the effects of training epochs on the network training and validation are observed more clearly in the minimum SSIM of the predictions. The low training SSIM can be associated with outliers in the training data set, which are captured accurately by the network only once it is trained sufficiently. The mean training SSIM, mean validation SSIM, minimum training SSIM and minimum validation SSIM, are 0.98, 0.985, 0.82 and 0.952, respectively.

\begin{figure}[ht]
\centering
    \begin{subfigure}[b]{.48\textwidth}
        \centering
        \includegraphics[width=\textwidth]{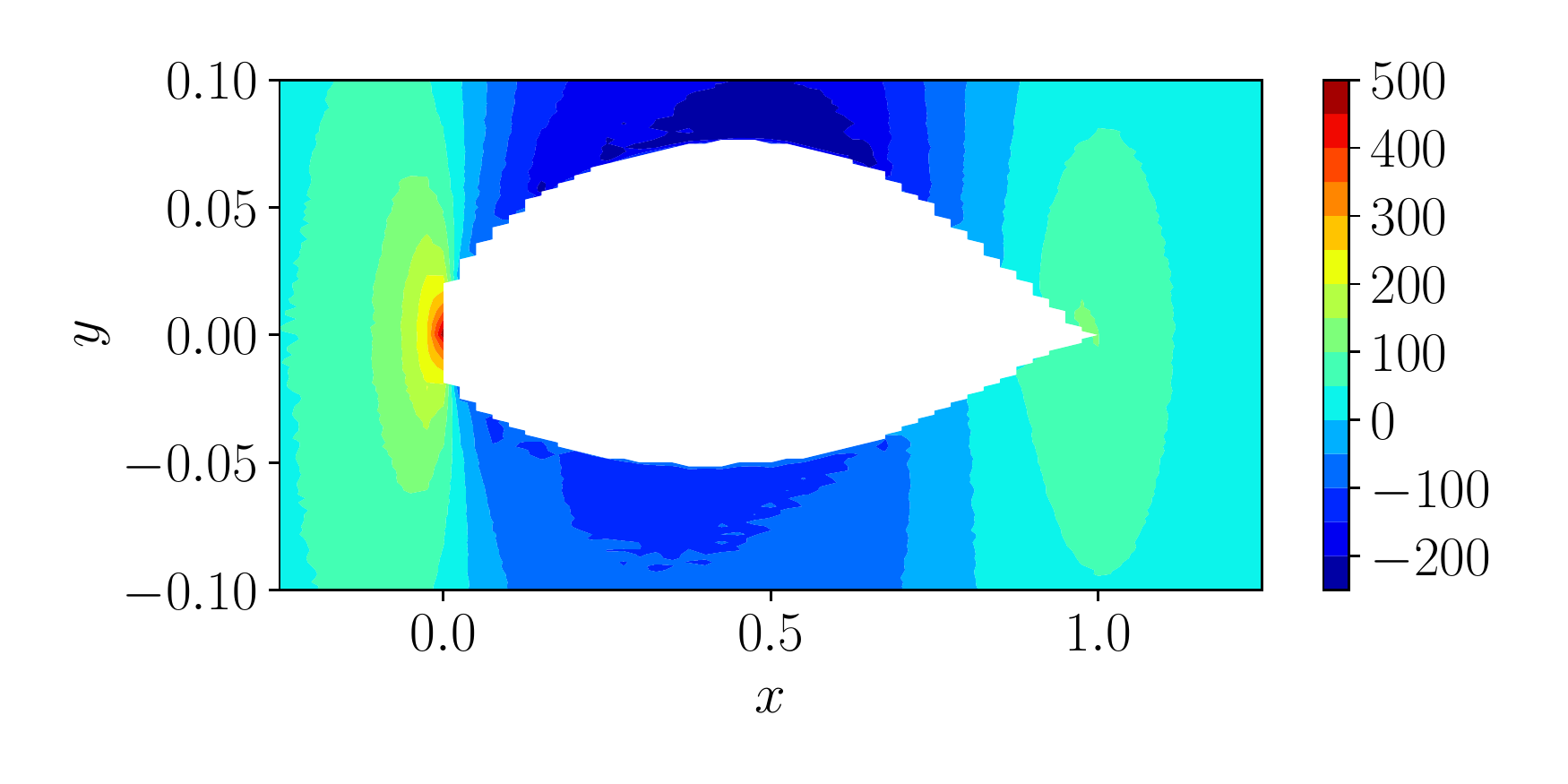}
        \caption{Convolution encoder-decoder prediction}
    \end{subfigure}
    \hfill
    \begin{subfigure}[b]{.48\textwidth}
        \centering
        \includegraphics[width=\textwidth]{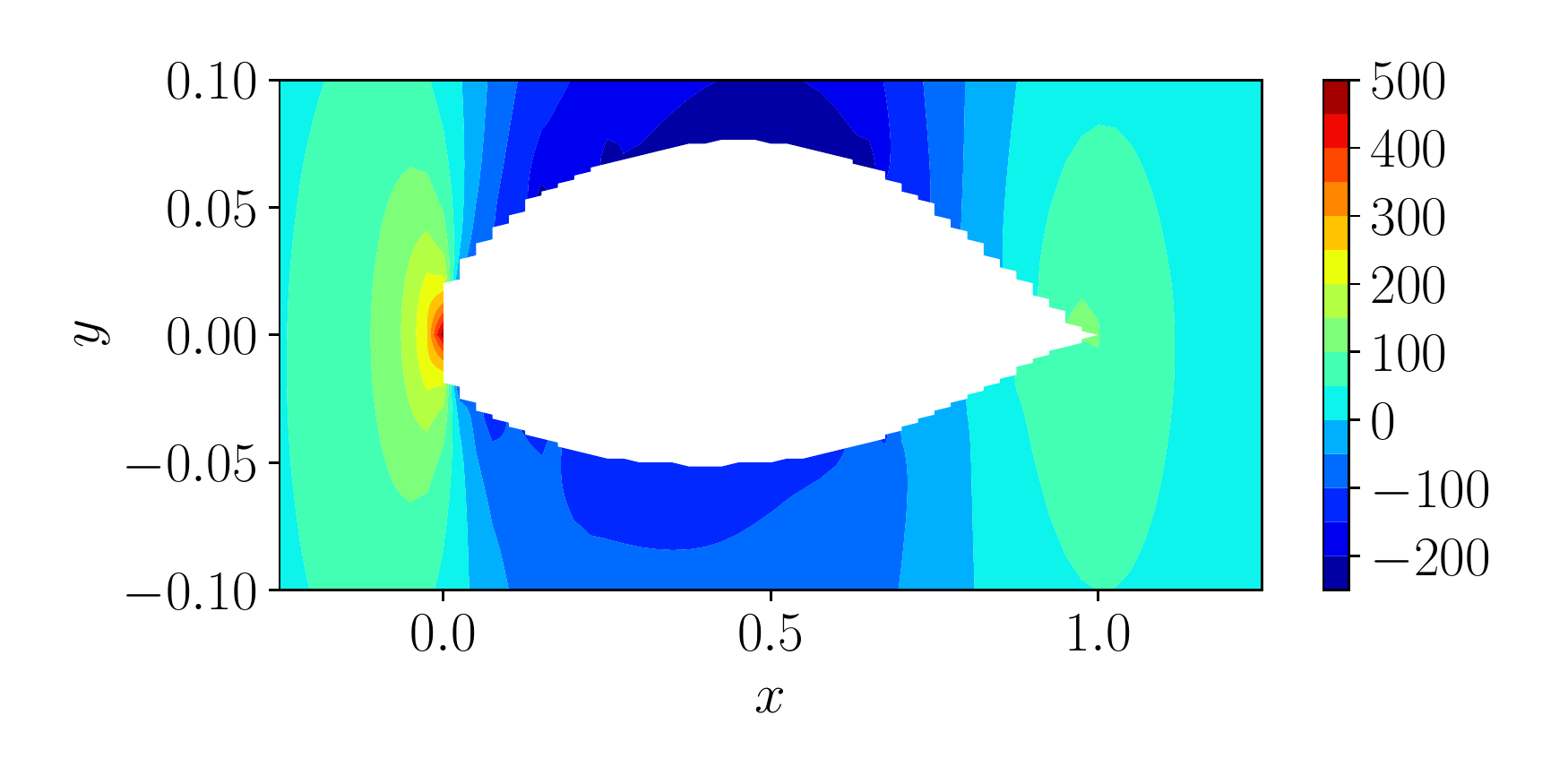}
        \caption{True solution}
    \end{subfigure}
\caption{Convolutional encoder-decoder prediction (a) and the true solution(b), for representative validation case}
\label{fig:CNN-enc-dec_p_pred}
\end{figure}

The convolutional encoder-decoder prediction and the true solution are shown in Figs. \ref{fig:CNN-enc-dec_p_pred} (a) and (b), respectively, for a representative validation case. As we can see the predicted solution matches the true solution very closely over most of the domain. The largest differences in the pressure are observed near the solid interface, on the top and bottom surfaces. However, the magnitude of the differences is small compared to the true pressure values in this region. This representative validation case has an SSIM of 0.985, which is almost the same as the mean validation SSIM. Thus, we can see that the convolutional encoder-decoder network has learned the pressure flow field around various hydrofoil surfaces and can perform generalized prediction for out-of-training cases.

\subsection{Shape optimization results}
Here we will present the results of morphing the NACA66 hydrofoil into minimum drag configurations subject to a thickness constraint and various lift constraints. Each design lift constraint represents a different optimization problem. We select $c_l=0.115 \pm 0.005$ and $c_l=0.147 \pm 0.006$, representing a lower design lift condition and a higher design lift condition, respectively, compared to the lift coefficient of 0.132 of the NACA66 at zero degrees. A tolerance is applied to each design lift for relaxing the constraint and posing the constraint as an inequality constraint. Thus, we want to morph the NACA66 configuration into a minimum drag configuration for a lower and a higher design condition. For both lift constraints, a thickness constraint of 11\% was selected. The optimization was initiated by selecting a perturbation variable set from the design variable bounds via random sampling and obtaining an initial perturbed configuration of the NACA66. A gradient-based optimization was then performed via MATLAB's Sequential Quadratic Programming (SQP) algorithm \cite{matlabopt}. 

\begin{figure}[ht]
\centering
\includegraphics[width=0.96\textwidth]{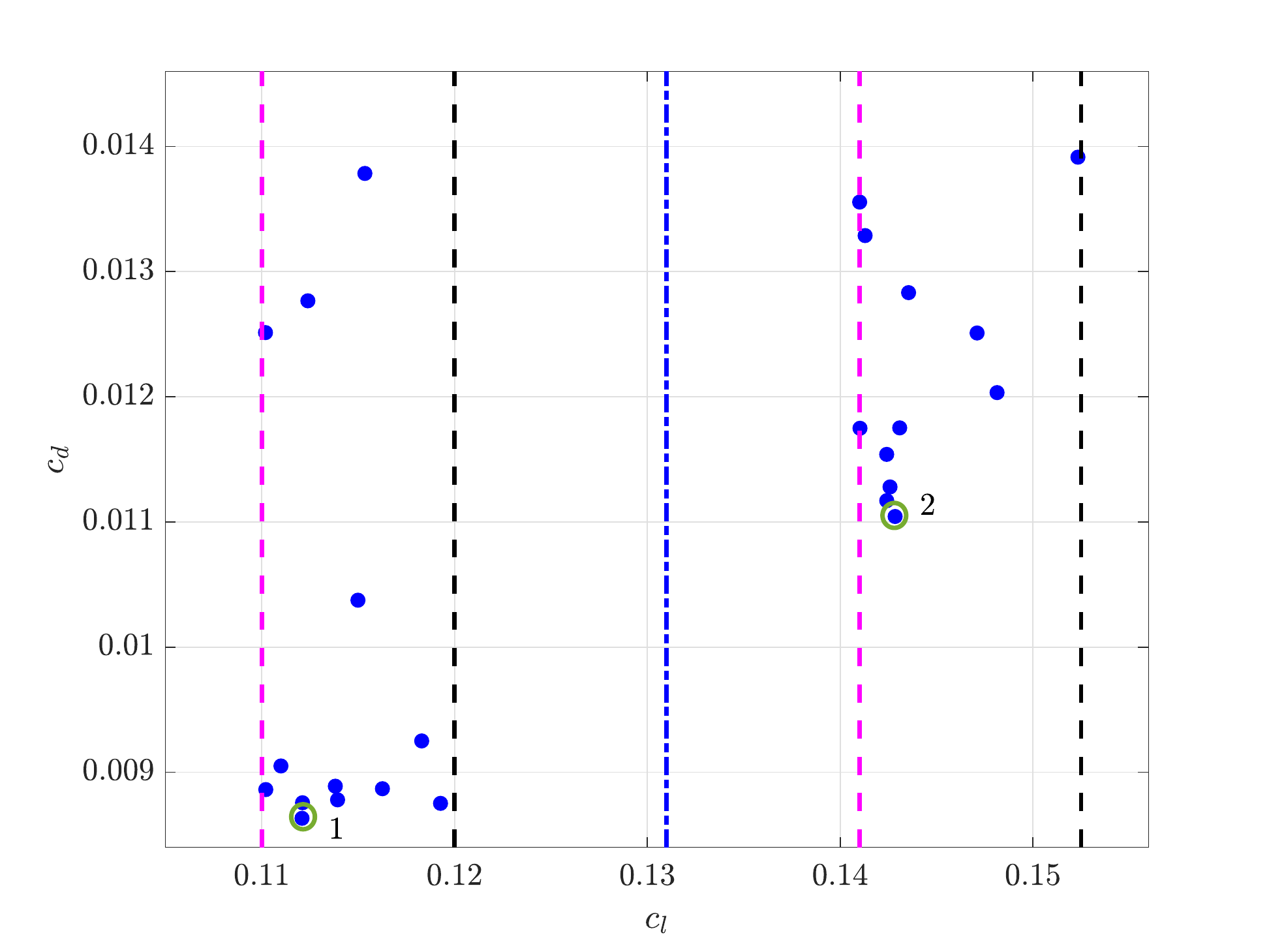}
\caption{Shape optimization results for $c_l$ constraints $c_l=0.115 \pm 0.005$ (1) and $c_l=0.147 \pm 0.006$ (2), with the best design shown for each constraint. Magenta and black represent the lower and upper bound of the constraint, respectively, for each design $c_l$ case. The blue dotted line represents the NACA66 $c_l$ as a reference.}
\label{fig:opt_results}
\end{figure}

Multimodality and the presence of local optima can be expected in both aerodynamic and hydrodynamic design \cite{chernukhin2013multimodality}. Thus, we perform the gradient-based optimization for 25 different initial perturbation sets for each of the design lift conditions. These 25 initial design variable sets were obtained from the design variable bounds via a Latin Hypercube Sampling. The results of the shape optimization for the different $c_l$ cases are shown in Fig. \ref{fig:opt_results}, where magenta and black represent the lower and upper bound of the constraints, respectively, for each design $c_l$. Only the feasible designs for each optimization case are shown and the best design for each $c_l$ constraint band ($c_l=0.115 \pm 0.005$ (1) and $c_l=0.147 \pm 0.006$ (2)) are encircled in green. The $c_l$ of NACA66 at zero degrees angle of attack is shown in blue as a reference. The results clearly indicate the presence of several local optima in the design space for each lift constraint band. A more comprehensive exploration of the design space study could be performed in the future with metaheuristic global optimization algorithms \cite{samanipour2020adaptive} to obtain the global optimum.

\begin{figure}[ht]
\centering
\includegraphics[width=0.96\textwidth]{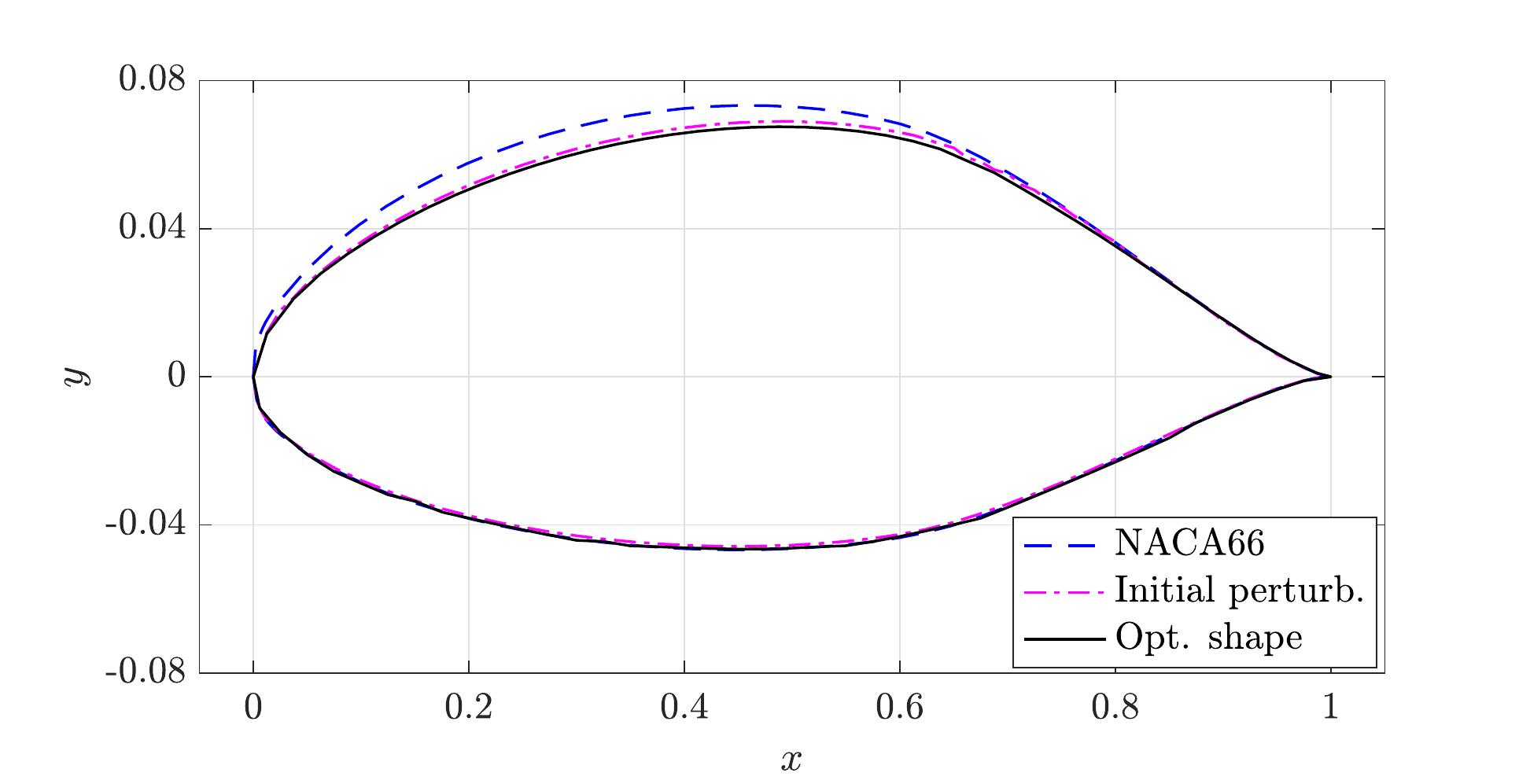}
\caption{Comparison of NACA66, initial perturbation and converged optimized shape for design 1}
\label{fig:comp_des1}
\end{figure}

\begin{figure}[ht]
\centering
\includegraphics[width=0.96\textwidth]{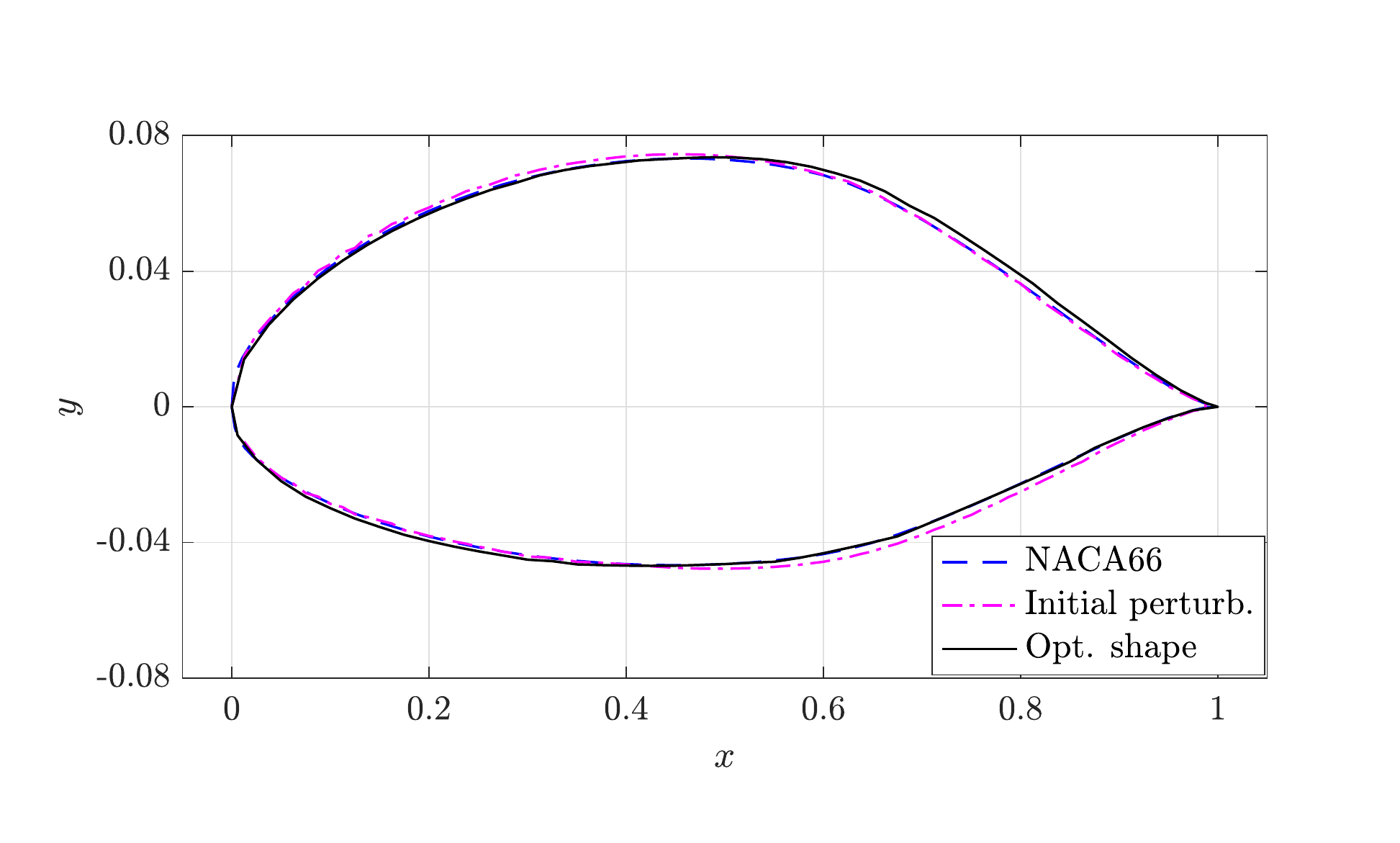}
\caption{Comparison of NACA66, initial perturbation and converged optimized shape for design 2}
\label{fig:comp_des2}
\end{figure}

The initial perturbed shape and the final optimized configuration for designs 1 and 2 are shown in Figs. \ref{fig:comp_des1} and \ref{fig:comp_des2}, respectively. The NACA66 hydrofoil is also presented as a reference. For the low $c_l$ the best local optimum (design 1), the initial perturbed shape is somewhat thinner than the NACA66. The final optimized design converged very close to the initial shape, which is corroborated by the 4 iterations required for optimization convergence. For the best local optimum (design 2) with a higher $c_l$, the initial shape and optimized shape show a greater difference, especially in the trailing edge of the lower surface. Compared to the NACA66 hydrofoil, the optimized shape for a higher design $c_l$ is slightly thicker near the leading edge of the bottom surface and trailing edge of the top surface.

Next, we will assess the accuracy of the best local optimum via the convolutional encoder-decoder-based optimization. Such an assessment is performed by validating the flow field and force predictions of a few representative optimized designs obtained from the surrogate-based optimization with full-order CFD simulations. 

\begin{figure}[ht]
\centering
    \begin{subfigure}[b]{.48\textwidth}
        \centering
        \includegraphics[width=\textwidth]{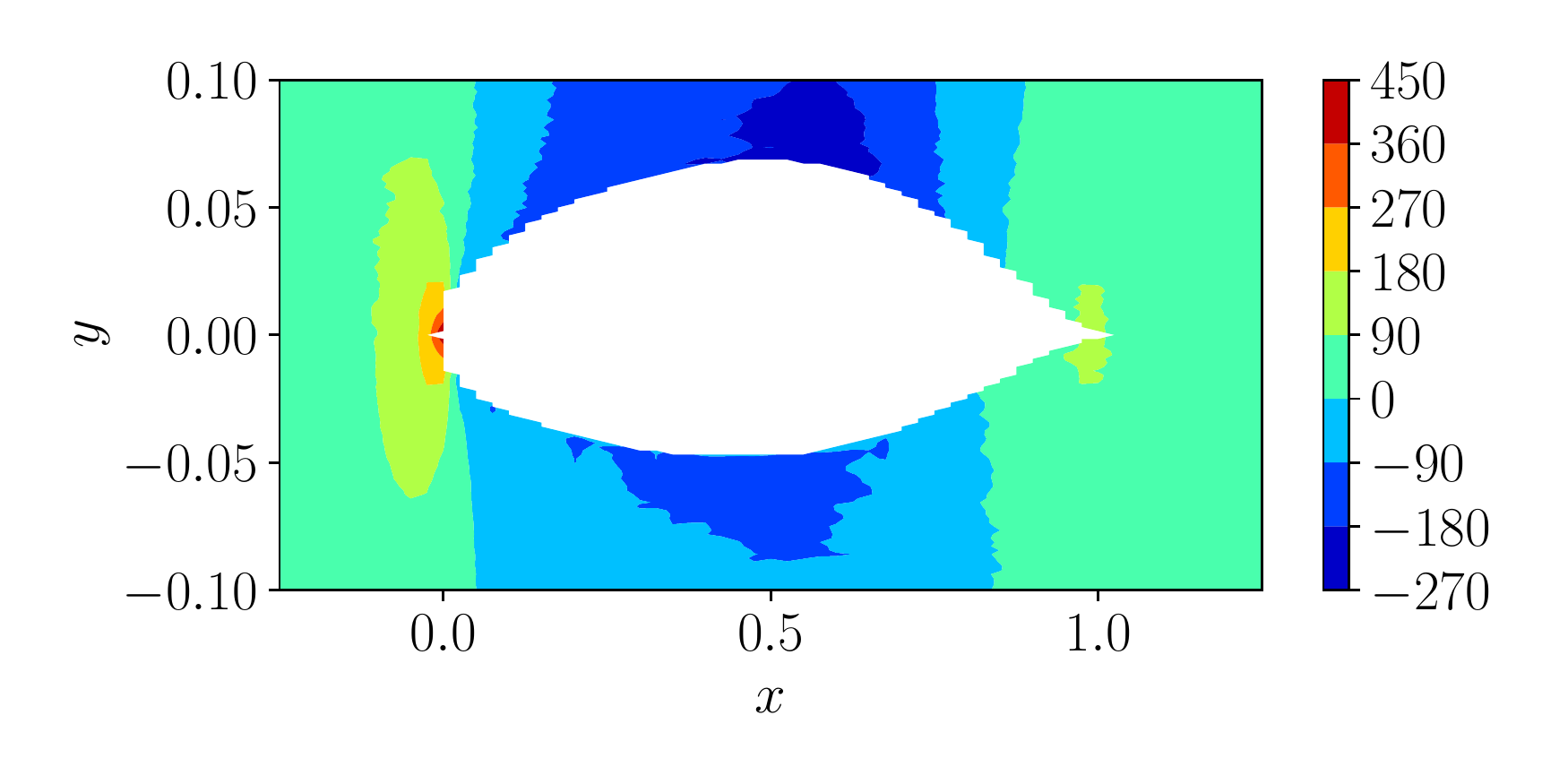}
        \caption{Predicted solution}
    \end{subfigure}
    \hfill
    \begin{subfigure}[b]{.48\textwidth}
        \centering
        \includegraphics[width=\textwidth]{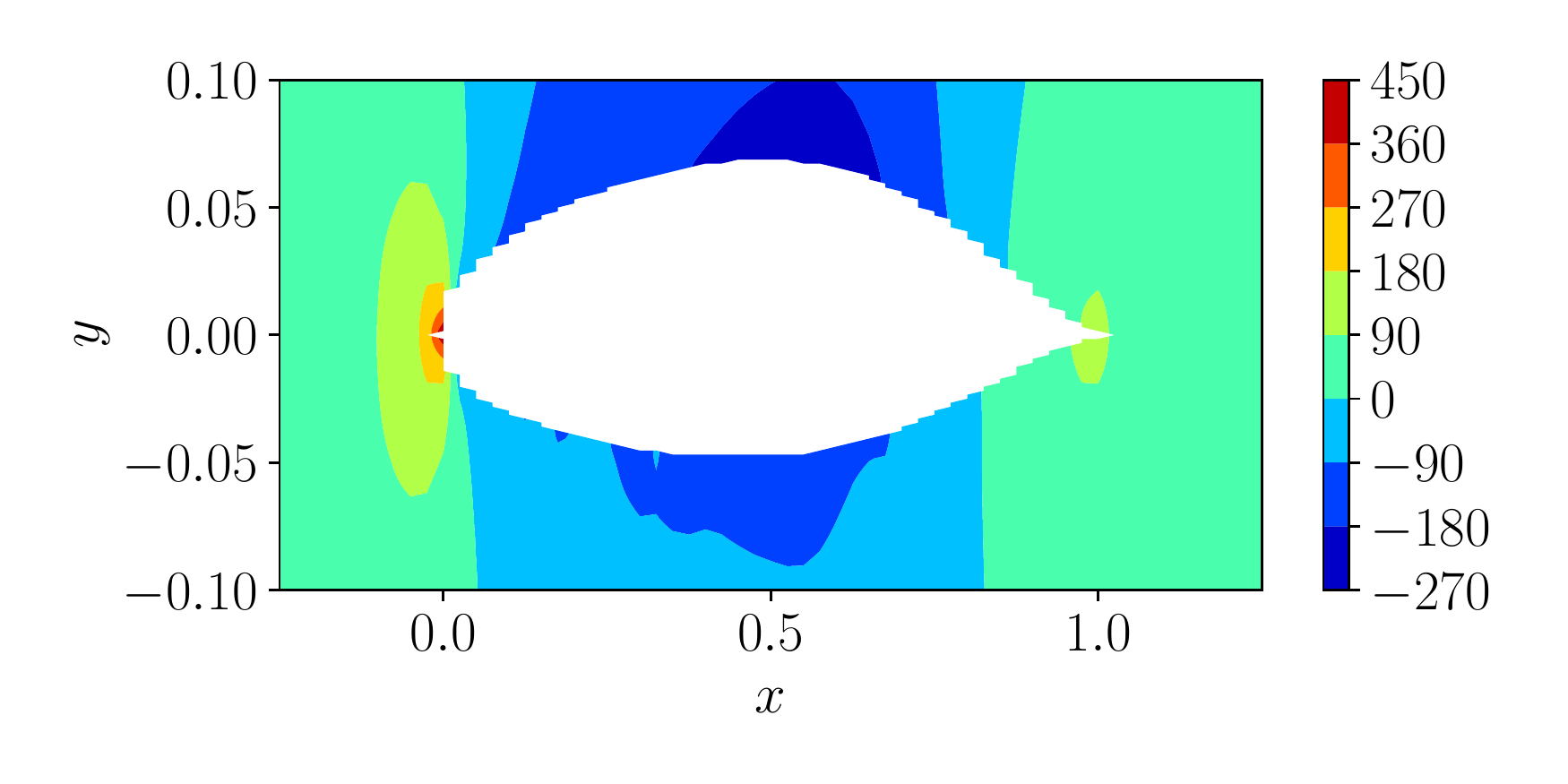}
        \caption{True solution}
    \end{subfigure}
\caption{Convolutional encoder-decoder prediction (a), and the true solution (b), for design 1}
\label{fig:comp_true_ped_des1}
\end{figure}

\begin{figure}[ht]
\centering
    \begin{subfigure}[b]{.48\textwidth}
        \centering
        \includegraphics[width=\textwidth]{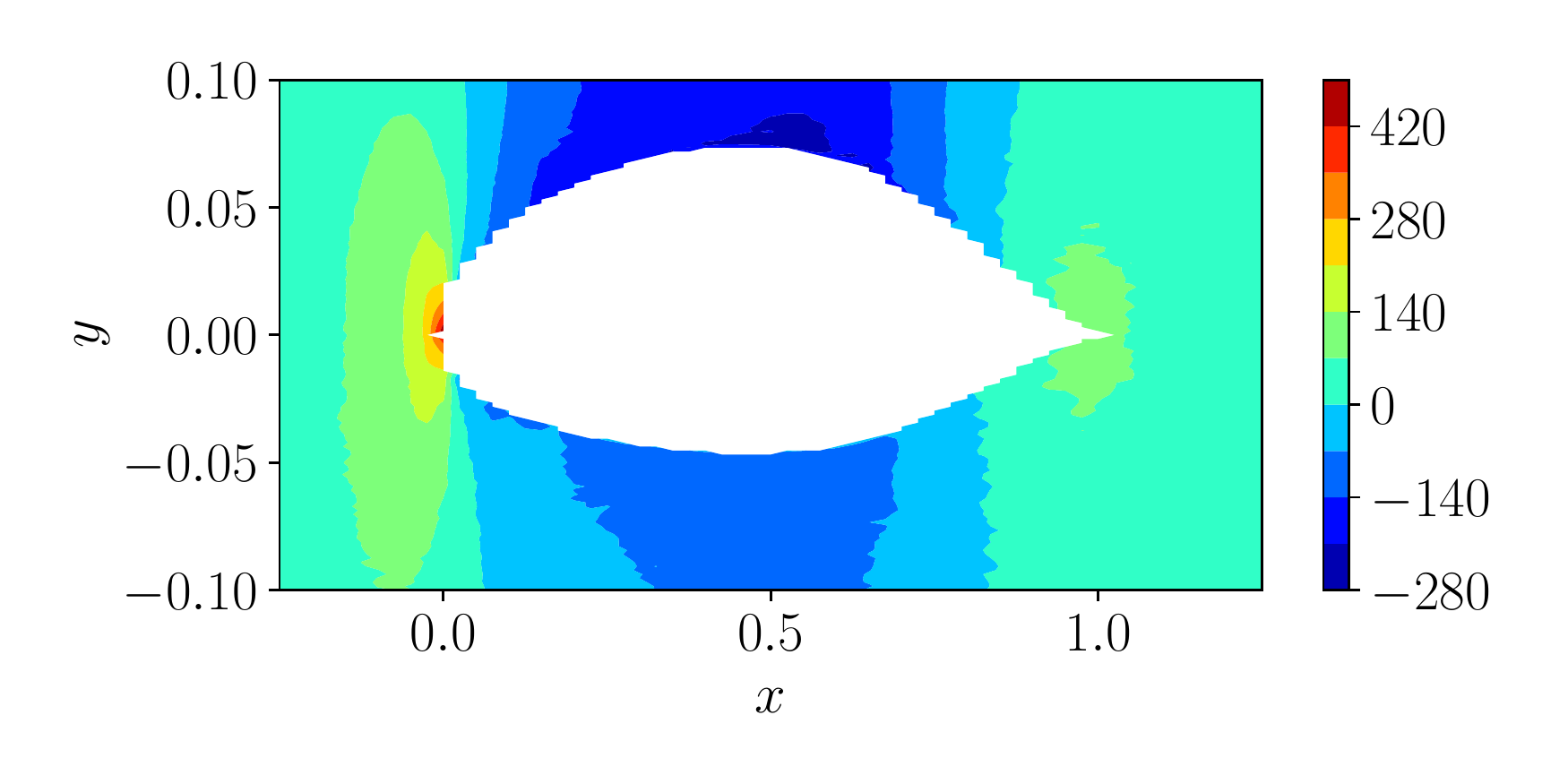}
        \caption{Predicted solution}
    \end{subfigure}
    \hfill
    \begin{subfigure}[b]{.48\textwidth}
        \centering
        \includegraphics[width=\textwidth]{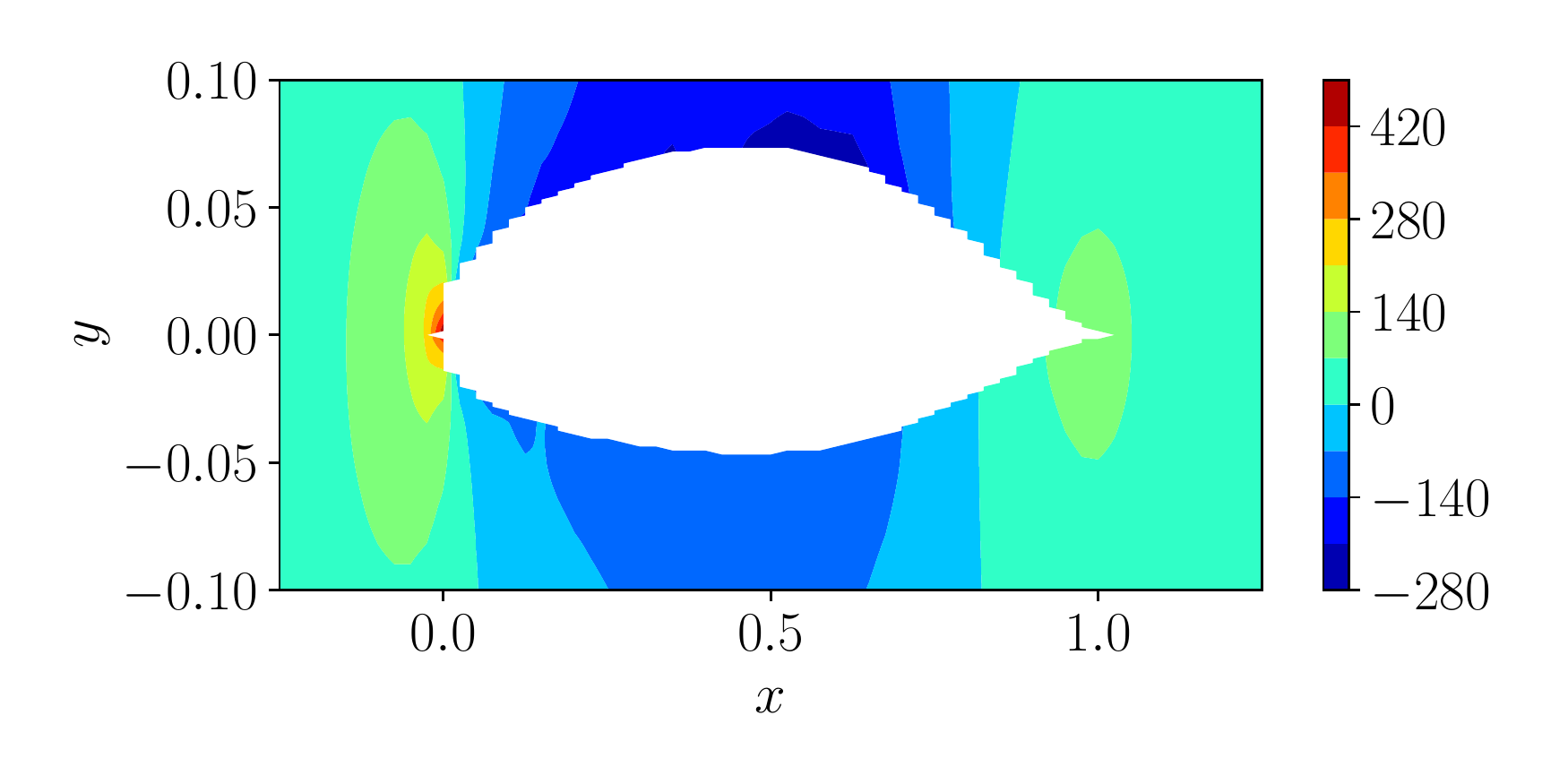}
        \caption{True solution}
    \end{subfigure}
\caption{Convolutional encoder-decoder prediction (a), and the true solution (b), for design 2}
\label{fig:comp_true_ped_des2}
\end{figure}

The pressure flow field predicted by the convolutional encoder-decoder around design 1 is compared to its counterpart obtained from full-order CFD simulation in Figs. \ref{fig:comp_true_ped_des1} (a) and (b), respectively. A similar comparison is presented for design 2 in Fig. \ref{fig:comp_true_ped_des2}. We can see that for design 1 the predicted solutions show some differences in the top surface suction and in the flow field away from the hydrofoil interface. For design 2, the top surface suction is predicted more accurately. Some differences in the bottom surface suction are also observed for both designs but their location varies for designs 1 and 2. Overall, the predicted flow reasonably matches the true flow field, which is corroborated by the SSIM of 0.95 and 0.98 for designs 1 and 2, respectively. The accuracy of the predicted flow fields by the convolutional encoder-decoder online application while being trained on a moderately large number of training cases shows its ability for generalized learning of flow physics. The lift and drag retrieved from the predicted hydrofoils were also compared to their counterparts obtained from the true solutions. For design 1, lift and drag coefficients obtained from the convolutional encoder-decoder predictions show a 3\% and 11\% absolute difference, respectively, compared to the full-order predictions. For design 2 the absolute difference in the lift and drag coefficients are 1\% and 12\%, respectively.

We can see that even though the predicted match the true solutions reasonably well, there is always a difference of around 10\% in the drag during the force retrieval from the pressure field. This remains consistent with both the validation case as well as designs 1 and 2. Since the ranking of the optimized designs depends on the actual drag coefficient values, we would also like to inspect the accuracy of the best local optimum obtained via surrogate-based optimization. To this end, the six best local optima were selected for each $c_l$ constraint, and their lift and drag predictions were compared to the full-order model predictions. The comparison of the retrieved lift and drag coefficients for $c_l^*=0.115 \pm 0.005$ are compared to the CFD predictions in Table \ref{tab:sixbestlocalopt_lowcl}. We can see that the high-order simulations also confirm design 1 as the lowest drag configuration. However, the fifth-ranked local optimum would violate the lift constraint according to the CFD predictions. Overall, the mean absolute difference of the lift and drag coefficients from encoder-decoder-predicted flow was $3\%$ and $11\%$, respectively, compared to the full-order predictions for the low $c_l$ case.
\begin{table}[ht]
\caption{Comparison of lift and drag coefficients of the six best local optima with full-order model predictions for $c_l^* = 0.115 \pm + 0.005$}
\centering
\begin{center}
\begin{tabular}{|l|c|c|c|c|}\hline
    &\multicolumn{2}{c|}{$c_l$} & \multicolumn{2}{c|}{$c_d$}\\
    \hline
    Rank & Retrieved & True & Retrieved & True\\
    \hline
    1 & 1.12e-1 & 1.16e-1 & 8.61e-3 & 9.72e-3\\
    2 & 1.19e-1 & 1.17e-1 & 8.75e-3 & 9.78e-3\\
    3 & 1.12e-1 & 1.18e-1 & 8.76e-3 & 1.02e-2\\
    4 & 1.14e-1 & 1.15e-1 & 8.78e-3 & 9.77e-3    \\
    5 & 1.10e-1 & \mymagenta{1.07e-1} & 8.86e-3 & 9.96e-3    \\   
    6 & 1.16e-1 & 1.10e-1 & 8.87e-3 & 1.00e-2\\
    \hline
\hline
\end{tabular}
\end{center}
\label{tab:sixbestlocalopt_lowcl}
\end{table}

The comparison of the retrieved lift and drag coefficients for $c_l^*=0.147 \pm 0.006$ are compared to the CFD predictions in Table \ref{tab:sixbestlocalopt_highcl}. For the higher $c_l$ constraint, the second-ranked local optimum turned out to be the lowest drag configuration according to the full-order predictions. The true drag predictions for all the optima were quite close to one another. However, the second-ranked optimum violated the $c_l$ constraint. Overall, the mean absolute difference of the lift and drag coefficients from encoder-decoder-predicted flow was $3.3\%$ and $13\%$, respectively. Thus, the optimization can lead to either a best local optimum or to the neighbourhood of the best local optimum for both the design $c_l$ cases. However, a higher-fidelity inspection is suggested to validate or locate the best local optimum.
\begin{table}[ht]
\caption{Comparison of lift and drag coefficients of the six best local optima with full-order model predictions for $c_l^* = 0.147 \pm + 0.006$}
\centering
\begin{center}
\begin{tabular}{|l|c|c|c|c|}\hline
    &\multicolumn{2}{c|}{$c_l$} & \multicolumn{2}{c|}{$c_d$}\\
    \hline
    Rank & Retrieved & True & Retrieved & True\\
    \hline
    1 & 1.42e-1 & 1.40e-1 & 1.15e-2 & 1.02e-2\\
    2 & 1.42e-1 & \mymagenta{1.37e-1} & 1.16e-2 & 1.04e-2\\
    3 & 1.43e-1 & 1.44e-1 & 1.17e-2 & 1.03e-2 \\
    4 & 1.41e-1 & 1.51e-1 & 1.18e-2 & 1.01e-2\\
    5 & 1.43e-1 & 1.44e-1 & 1.18e-2 & 1.03e-2\\
    6 & 1.48e-1 & 1.56e-1 & 1.20e-2 & 1.06e-2
    \\
    \hline
\hline
\end{tabular}
\end{center}
\label{tab:sixbestlocalopt_highcl}
\end{table}


The online flow-field prediction via the convolutional encoder-decoder takes less than ten seconds on a 2.1 Ghz Intel E5-2683 v4 Broadwell CPU. On the other hand, each RANS requires close to 2.8e5 CPU seconds on the same computational facility, which is almost five orders higher computational cost compared to the convolutional encoder-decoder prediction. The training cost of the convolutional encoder-decoder network can thus be recovered quickly once the surrogate model is involved in hundreds of flow field computations during the optimization. 

It is obvious that employing only six design variables significantly reduced the shape sensitivity computations and enabled us to perform real-time optimizations without resorting to adjoint-based design optimization techniques. Comparing the true computational efficiency of the present parametric level set-based approach over routine FFD-based optimization can be performed in a future study.

\section{Conclusions}
In this article, we have presented a new approach of implicitly morphing hydrofoil shapes during shape optimization via perturbation of parametric level sets. The morphing approach consists of perturbing the polynomial coefficients of the RBF-based parametric level sets with only six parameters. Thus, with these six parameters set as design variables, we can morph baseline hydrofoil configurations into various other hydrofoil shapes via a shape optimization algorithm. Most importantly, the employment of only six shape design variables is not only lower than the number of shape design variables required in the FFD-based shape optimization but several orders lower than the conventional application of RBF-based parametric level sets for topology optimization. This makes the present parametric level set approach a computationally efficient avenue for shape optimization task. Furthermore, the fixed uniform Cartesian level set mesh structure also facilitates the application of convolutional encoder-decoder networks as a surrogate of high-fidelity RANS simulations for the flow field prediction around morphed hydofoil boundaries. Such an application of shape morphing via a parametric level set method, which can also integrate a CNN-based surrogate model with the optimization, is presented here for the first time. 

The results presented here show that with optimal training the convolutional encoder-decoder can perform generalized prediction of pressure field around various hydrofoil configurations, even when a moderate number of training samples were considered. This is indicated by the mean SSIM of 0.985 and a minimum SSIM of 0.95 for predicted flow fields compared to the true solutions, for a range of out-of-training hydrofoil configurations. While the convolutional encoder-decoder networks have been employed to predict flow fields in the past, here we present a comprehensive demonstration of their generalized learning and prediction ability of flow fields over varying shapes. Most importantly, the convolutional encoder-decoder performs flow field prediction in real-time and almost five orders of magnitude faster compared to similar predictions via RANS. The convolutional encoder-decoder model was subsequently integrated with the shape optimization via the parametric level set-based morphing procedure. With the convolutional encoder-decoder surrogate model, a computationally tractable shape optimization of 25 initial hydrofoil shapes could be performed for each of the design $c_l$ constraints of $0.115 \pm 0.005$ and $0.147 \pm 0.006$. The results showed widely differing local optima for minimum drag configurations, which further corroborates the importance of adequate exploration of the design space to reach the global optimum.

To assess the accuracy of our convolutional encoder-decoder-based optimization we inspected the flow fields of the best local minimum drag for both the design $c_l$ cases with RANS predictions. The predicted flow fields showed an SSIM of 0.94 and 0.98, respectively, for the low and high $c_l$ cases, when compared to the true solutions. This indicated the accuracy of flow field predictions during the online optimization application of the convolutional encoder-decoder. Furthermore, the force coefficients for the six best local optima for each design $c_l$ case were compared with their corresponding CFD-based predictions. The lift and drag coefficient prediction errors were within 3\% and 12\%, respectively. Importantly, the surrogate-based best local optimum prediction either coincides with CFD-based prediction or lies in its close neighbourhood for both $c_l$ cases. This shows the promise of the present approach for future shape optimization over a broader spectrum of flow conditions and hydrofoil shapes. 

\section*{Acknowledgment}
This research was funded by the Natural Sciences and Engineering Research Council of Canada (NSERC) [grant number IRCPJ 550069-19]. The training and evaluation of the neural network models were supported in part by the computational resources and services provided by the Digital Research Alliance of Canada. 

\appendix


 \bibliographystyle{elsarticle-num} 
 \bibliography{cas-refs}

\begin{thebibliography}{10}
\expandafter\ifx\csname url\endcsname\relax
  \def\url#1{\texttt{#1}}\fi
\expandafter\ifx\csname urlprefix\endcsname\relax\def\urlprefix{URL }\fi
\expandafter\ifx\csname href\endcsname\relax
  \def\href#1#2{#2} \def\path#1{#1}\fi

\bibitem{haftka1986structural}
R.~T. Haftka, R.~V. Grandhi, Structural shape optimization—a survey, Computer
  methods in applied mechanics and engineering 57~(1) (1986) 91--106.

\bibitem{jameson_1998}
A.~Jameson, N.~Pierce, L.~Martinelli., Optimum aerodynamic design using the
  {N}avier-{S}tokes equations, J. Theor. Comp. Fluid Mech 10 (1998) 213--237.

\bibitem{Giles_2000}
M.~B. Giles, N.~A. Pierce, An introduction to the adjoint approach to design,
  Flow, Turbulence and Combustion 65 (2000) 393–415.

\bibitem{pironneau_2004}
B.~Mohammadi, O.~Pironneau, Shape optimization in fluid mechanics, Annual
  Review of Fluid Mechanics 36~(1) (2004) 255--279.

\bibitem{reuther1996aerodynamic}
J.~Reuther, A.~Jameson, J.~Farmer, L.~Martinelli, D.~Saunders, Aerodynamic
  shape optimization of complex aircraft configurations via an adjoint
  formulation, in: 34th Aerospace Sciences Meeting and Exhibit, 1996, p.~94.

\bibitem{samareh2004aerodynamic}
J.~Samareh, Aerodynamic shape optimization based on free-form deformation, in:
  10th AIAA/ISSMO multidisciplinary analysis and optimization conference, 2004,
  p. 4630.

\bibitem{garg2015high}
N.~Garg, G.~K. Kenway, Z.~Lyu, J.~R. Martins, Y.~L. Young, High-fidelity
  hydrodynamic shape optimization of a 3-d hydrofoil, Journal of Ship Research
  59~(4) (2015) 209--226.

\bibitem{lyu2015aerodynamic}
Z.~Lyu, G.~K. Kenway, J.~R. Martins, Aerodynamic shape optimization
  investigations of the common research model wing benchmark, AIAA journal
  53~(4) (2015) 968--985.

\bibitem{he2019robust}
X.~He, J.~Li, C.~A. Mader, A.~Yildirim, J.~R. Martins, Robust aerodynamic shape
  optimization—from a circle to an airfoil, Aerospace Science and Technology
  87 (2019) 48--61.

\bibitem{chernukhin2013multimodality}
O.~Chernukhin, D.~W. Zingg, Multimodality and global optimization in
  aerodynamic design, AIAA journal 51~(6) (2013) 1342--1354.

\bibitem{timme2011transonic}
S.~Timme, S.~Marques, K.~Badcock, Transonic aeroelastic stability analysis
  using a kriging-based schur complement formulation, AIAA journal 49~(6)
  (2011) 1202--1213.

\bibitem{fan2019reliability}
X.~Fan, P.~Wang, F.~Hao, Reliability-based design optimization of crane bridges
  using kriging-based surrogate models, Structural and Multidisciplinary
  Optimization 59~(3) (2019) 993--1005.

\bibitem{mallik2020kriging}
W.~Mallik, D.~E. Raveh, Kriging-based aeroelastic gust response analysis at
  high angles of attack, AIAA Journal 58~(9) (2020) 3777--3787.

\bibitem{raissi2016deep}
M.~Raissi, G.~Karniadakis, Deep multi-fidelity gaussian processes, arXiv
  preprint arXiv:1604.07484 (2016).

\bibitem{perdikaris2017nonlinear}
P.~Perdikaris, M.~Raissi, A.~Damianou, N.~D. Lawrence, G.~E. Karniadakis,
  Nonlinear information fusion algorithms for data-efficient multi-fidelity
  modelling, Proceedings of the Royal Society A: Mathematical, Physical and
  Engineering Sciences 473~(2198) (2017) 20160751.

\bibitem{keane2008engineering}
A.~Keane, A.~Forrester, A.~Sobester, Engineering design via surrogate
  modelling: a practical guide, American Institute of Aeronautics and
  Astronautics, Inc., 2008.

\bibitem{bonfiglio2018improving}
L.~Bonfiglio, P.~Perdikaris, G.~Vernengo, J.~S. de~Medeiros, G.~Karniadakis,
  Improving swath seakeeping performance using multi-fidelity gaussian process
  and bayesian optimization, Journal of Ship Research 62~(04) (2018) 223--240.

\bibitem{yao2020reduced}
W.~Yao, S.~Marques, T.~Robinson, C.~Armstrong, L.~Sun, A reduced-order model
  for gradient-based aerodynamic shape optimisation, Aerospace Science and
  Technology 106 (2020) 106120.

\bibitem{marques2021non}
S.~P. Marques, L.~Kob, T.~T. Robinson, W.~Yao, L.~Sun, Non-intrusive
  aerodynamic shape optimisation with a discrete empirical interpolation
  method, in: AIAA Scitech 2021 Forum, 2021, p. 0172.

\bibitem{bouhlel2020scalable}
M.~A. Bouhlel, S.~He, J.~R. Martins, Scalable gradient-enhanced artificial
  neural networks for airfoil shape design in the subsonic and transonic
  regimes, Structural and Multidisciplinary Optimization (2020) 1--14.

\bibitem{junior2022intelligent}
J.~M.~M. J{\'u}nior, G.~L. Halila, Y.~Kim, T.~Khamvilai, K.~G. Vamvoudakis,
  Intelligent data-driven aerodynamic analysis and optimization of morphing
  configurations, Aerospace Science and Technology 121 (2022) 107388.

\bibitem{bhatnagar2019prediction}
S.~Bhatnagar, Y.~Afshar, S.~Pan, K.~Duraisamy, S.~Kaushik, Prediction of
  aerodynamic flow fields using convolutional neural networks, Computational
  Mechanics 64~(2) (2019) 525--545.

\bibitem{xu2020multi}
J.~Xu, K.~Duraisamy, Multi-level convolutional autoencoder networks for
  parametric prediction of spatio-temporal dynamics, Computer Methods in
  Applied Mechanics and Engineering 372 (2020) 113379.

\bibitem{bronstein2017geometric}
M.~M. Bronstein, J.~Bruna, Y.~LeCun, A.~Szlam, P.~Vandergheynst, Geometric deep
  learning: going beyond euclidean data, IEEE Signal Processing Magazine 34~(4)
  (2017) 18--42.

\bibitem{mallik2022assessment}
W.~Mallik, R.~K. Jaiman, J.~Jelovica,
  \href{https://arxiv.org/abs/2204.05573}{Assessment of convolutional recurrent
  autoencoder network for learning wave propagation} (2022).
\newblock \href {https://doi.org/10.48550/ARXIV.2204.05573}
  {\path{doi:10.48550/ARXIV.2204.05573}}.
\newline\urlprefix\url{https://arxiv.org/abs/2204.05573}

\bibitem{sederberg1986free}
T.~W. Sederberg, S.~R. Parry, Free-form deformation of solid geometric models,
  in: Proceedings of the 13th annual conference on Computer graphics and
  interactive techniques, 1986, pp. 151--160.

\bibitem{kenway2010cad}
G.~Kenway, G.~Kennedy, J.~R. Martins, A {CAD}-free approach to high-fidelity
  aerostructural optimization, in: 13th AIAA/ISSMO multidisciplinary analysis
  optimization conference, 2010, p. 9231.

\bibitem{mallik2022deep}
W.~Mallik, N.~Farvolden, R.~K. Jaiman, J.~Jelovica, Deep convolutional neural
  network for shape optimization using level-set approach, arXiv preprint
  arXiv:2201.06210 (2022).

\bibitem{sethian1996theory}
J.~A. Sethian, Theory, algorithms, and applications of level set methods for
  propagating interfaces, Acta numerica 5 (1996) 309--395.

\bibitem{allaire2002level}
G.~Allaire, F.~Jouve, A.-M. Toader, A level-set method for shape optimization,
  Comptes Rendus Mathematique 334~(12) (2002) 1125--1130.

\bibitem{wang2006radial}
S.~Wang, M.~Y. Wang, Radial basis functions and level set method for structural
  topology optimization, International journal for numerical methods in
  engineering 65~(12) (2006) 2060--2090.

\bibitem{wang2007extended}
S.~Wang, K.~M. Lim, B.~C. Khoo, M.~Y. Wang, An extended level set method for
  shape and topology optimization, Journal of Computational Physics 221~(1)
  (2007) 395--421.

\bibitem{jiang2018parametric}
L.~Jiang, S.~Chen, X.~Jiao, Parametric shape and topology optimization: A new
  level set approach based on cardinal basis functions, International Journal
  for Numerical Methods in Engineering 114~(1) (2018) 66--87.

\bibitem{aghasi2011parametric}
A.~Aghasi, M.~Kilmer, E.~L. Miller, Parametric level set methods for inverse
  problems, SIAM Journal on Imaging Sciences 4~(2) (2011) 618--650.

\bibitem{pingen2010parametric}
G.~Pingen, M.~Waidmann, A.~Evgrafov, K.~Maute, A parametric level-set approach
  for topology optimization of flow domains, Structural and Multidisciplinary
  Optimization 41~(1) (2010) 117--131.

\bibitem{guirguis2018high}
D.~Guirguis, W.~W. Melek, M.~F. Aly, High-resolution non-gradient topology
  optimization, Journal of Computational Physics 372 (2018) 107--125.

\bibitem{vernengo2016physics}
G.~Vernengo, L.~Bonfiglio, S.~Gaggero, S.~Brizzolara, Physics-based design by
  optimization of unconventional supercavitating hydrofoils, Journal of Ship
  Research 60~(04) (2016) 187--202.

\bibitem{gaggero2017efficient}
S.~Gaggero, G.~Tani, D.~Villa, M.~Viviani, P.~Ausonio, P.~Travi, G.~Bizzarri,
  F.~Serra, Efficient and multi-objective cavitating propeller optimization: An
  application to a high-speed craft, Applied Ocean Research 64 (2017) 31--57.

\bibitem{miglianti2020predicting}
L.~Miglianti, F.~Cipollini, L.~Oneto, G.~Tani, S.~Gaggero, A.~Coraddu,
  M.~Viviani, Predicting the cavitating marine propeller noise at design stage:
  A deep learning based approach, Ocean Engineering 209 (2020) 107481.

\bibitem{sethian1999level}
J.~A. Sethian, Level set methods and fast marching methods: evolving interfaces
  in computational geometry, fluid mechanics, computer vision, and materials
  science, Vol.~3, Cambridge university press, 1999.

\bibitem{wendland1995piecewise}
H.~Wendland, Piecewise polynomial, positive definite and compactly supported
  radial functions of minimal degree, Advances in computational Mathematics
  4~(1) (1995) 389--396.

\bibitem{martins2013review}
J.~R. Martins, J.~T. Hwang, Review and unification of methods for computing
  derivatives of multidisciplinary computational models, AIAA Journal 51~(11)
  (2013) 2582--2599.

\bibitem{deb2002fast}
K.~Deb, A.~Pratap, S.~Agarwal, T.~Meyarivan, A fast and elitist multiobjective
  genetic algorithm: Nsga-ii, IEEE transactions on evolutionary computation
  6~(2) (2002) 182--197.

\bibitem{bukka2021assessment}
S.~R. Bukka, R.~Gupta, A.~R. Magee, R.~K. Jaiman, Assessment of unsteady flow
  predictions using hybrid deep learning based reduced-order models, Physics of
  Fluids 33~(1) (2021) 013601.

\bibitem{joshi2017variationally}
V.~Joshi, R.~K. Jaiman, A variationally bounded scheme for delayed detached
  eddy simulation: application to vortex-induced vibration of offshore riser,
  Computers \& fluids 157 (2017) 84--111.

\bibitem{kashyap2021robust}
S.~R. Kashyap, R.~K. Jaiman, A robust and accurate finite element framework for
  cavitating flows with moving fluid-structure interfaces, Computers \&
  Mathematics with Applications 103 (2021) 19--39.

\bibitem{kingma2014adam}
D.~P. Kingma, J.~Ba, Adam: A method for stochastic optimization, arXiv preprint
  arXiv:1412.6980 (2014).

\bibitem{matlabdl}
I.~The~MathWorks, \href{https://www.mathworks.com/help/deeplearning/}{Matlab
  deep learning toolbox} (2022).
\newline\urlprefix\url{https://www.mathworks.com/help/deeplearning/}

\bibitem{wang2004image}
Z.~Wang, A.~C. Bovik, H.~R. Sheikh, E.~P. Simoncelli, Image quality assessment:
  from error visibility to structural similarity, IEEE transactions on image
  processing 13~(4) (2004) 600--612.

\bibitem{mallik2022predicting}
W.~Mallik, R.~K. Jaiman, J.~Jelovica, Predicting transmission loss in
  underwater acoustics using convolutional recurrent autoencoder network, The
  Journal of the Acoustical Society of America 152~(3) (2022) 1627--1638.

\bibitem{matlabopt}
I.~The~MathWorks, \href{https://www.mathworks.com/help/optim/}{Matlab
  optimization toolbox} (2022).
\newline\urlprefix\url{https://www.mathworks.com/help/optim/}

\bibitem{samanipour2020adaptive}
F.~Samanipour, J.~Jelovica, Adaptive repair method for constraint handling in
  multi-objective genetic algorithm based on relationship between constraints
  and variables, Applied Soft Computing 90 (2020) 106143.

\end{thebibliography}





\end{document}